\documentclass[12pt]{article}
\usepackage{amsmath, amssymb, graphics, amsfonts}

\usepackage{feynmp}
\usepackage{amsbsy}
\usepackage{amsfonts}
\usepackage{amsmath}
\usepackage{graphicx}

\newcommand{\mathsym}[1]{{}}

\usepackage[mathscr]{eucal}

\usepackage{latexsym, graphics, makeidx, epsfig}

\newlength{\abstractwidth}

\def\id{\protect{{1 \kern-.28em {\rm l}}}}

\def \ci  {\cite}
\def \foot{\footnote}

\def\p{{\partial}}

\makeatletter
\renewcommand\section{\@startsection {section}{1}{\z@}%
                                   {-3.5ex \@plus -1ex \@minus -.2ex}%
                                   {2.3ex \@plus.2ex}%
                                   {\normalfont\large\bfseries}}
\renewcommand\subsection{\@startsection{subsection}{2}{\z@}%
                                   {-3.25ex\@plus -1ex \@minus -.2ex}%
                                   {1.5ex \@plus .2ex}%
                                   {\normalfont\normalsize\bfseries}}
\makeatother

\def\d{{\rm d}} 

\def\Tr{{\rm tr}}

\def\no{\nonumber}

\def\rC{{\rm C}}

\def\rC{ V}

\def \vp {\varphi}

\numberwithin{equation}{section} \makeatletter
\@addtoreset{equation}{section}

\vspace{-1cm}

  \def \te {\theta}
  \def \vp {\varphi}

\newcommand{\be}{\begin{eqnarray}}
\newcommand{\ee}{\end{eqnarray}}
\newcommand{\bea}{\begin{eqnarray}}
\newcommand{\eea}{\end{eqnarray}}

\newcommand{\rf}[1]{(\ref{#1})}

\def \lam {{\ell}}

\def\la{\label}
\def\ov{\over}
\def\ha{{ 1\over 2}}
\def \te {\textstyle} 

\def\ts{{\tilde j}}

\def\Tr{{\rm Tr}}

\def\Tr{{\rm Tr}}

\def\spa#1.#2{\left\langle#1\,#2\right\rangle}
\def\spb#1.#2{\left[#1\,#2\right]}
\def\sandmx#1.#2.#3{%
\left\langle#1{\vphantom1}\right|{#2}\left|#3\right]}%

\def\Tr{{\rm Tr}}
\def\ts{{\tilde j}}
\def\tsj{j'}
\def \iffa {\iffalse}
\def \fo {{1\ov 4}}\def \no {\nonumber}

\def\pic #1#2#3{\hbox{\lower#1pt\hbox{~\mbox{\includegraphics[height=#3mm]{#2}}}}}

\def \ts {{j'}}
\def \ha  {{\textstyle{{1\ov 2}}}}
\def \half {{1\ov 2}}
\def \k {\kappa}
\def \yg {Y}
\def \cC  {{\rm C}}
 \def \vp {\varphi}
 \def \l {\lambda} 
\def \bi {\bibitem}
\def \wC {{\cal V}}
\def \cA   {{\cal A}}

\begin{document}

\date{\currenttime}
\today
\date {}
\vspace{1 cm}

\overfullrule=0pt
\parskip=2pt
\parindent=12pt
\headheight=0in \headsep=0in \topmargin=0in \oddsidemargin=0in

\thispagestyle{empty}
\vspace{-1cm}

\rightline{Imperial-TP-AAT-2017-01}


\begin{center}
\vspace{1cm}
{\Large\bf  
 On  four-point interactions in  massless
 
 \vspace{0.2cm}
  higher spin theory in flat space 
 }
\vspace{1.0cm}

\vspace{.2cm}
 { R. Roiban$^{a}$    and 
 A.A. Tseytlin$^{b,}$\footnote{Also at Lebedev  Institute, Moscow. }
}

\vskip 0.2cm

{
\em 
\vskip 0.08cm
\vskip 0.08cm 
$^{a}$Department of Physics, The Pennsylvania  State University,\\
University Park, PA 16802 , USA\\
\vskip 0.08cm
\vskip 0.08cm 
$^{b}$Blackett Laboratory, Imperial College,
London SW7 2AZ, U.K.
 }
\vspace{.2cm}
\end{center}

\newcommand{\ar}[1]{{  {#1}  }}

\begin{abstract}
 We consider a  minimal  interacting theory of a single tower of  spin $j=0,2,4,...$  massless  
 Fronsdal fields in flat space  \ar{with local
 Lorentz}-covariant  cubic  interaction vertices. 
  We address the question  of constraints  on  possible 
  quartic   interaction vertices imposed  by the condition of  on-shell gauge 
  invariance of the tree-level four-point scattering amplitudes  involving three   spin 0 and one spin $j$  particle. 
   We find that  these constraints admit a local solution for 
  quartic $000j$ interaction term in the action only  for $j=2$ and $j=4$. 
  We determine 
  the
  non-local terms in four-vertices required in the $j \ge 6$ case   and 
  suggest    that these non-localities   
  may  be  interpreted  as a result of integrating out a tower of 
  auxiliary   ghost-like 
  massless higher spin fields   in an extended  theory with a local action, up 
to possible higher-point interactions of the ghost fields. 
 We   also consider  the conformal off-shell  extension of the Einstein theory  and show that  the perturbative expansion 
 of its action 
 is the same as that of the non-local action resulting from integrating out the trace of the graviton field from the  Einstein 
action.
 Motivated by this  example, we  conjecture the  existence of a similar  
  conformal off-shell extension  of 
 a massless   higher spin theory   that   may be related to the above extended   theory. 
 It may then   have the same  infinite-dimensional 
 symmetry as  the higher-derivative  conformal higher spin theory  and    may thus  lead to  a  trivial S matrix.

\end{abstract}

\newpage 

{
\tableofcontents

}
\newpage
\setcounter{footnote}{0}


\section{Introduction}

The existence of  an 
 interacting theory   of massless   higher spins   in flat space
  is  usually  considered  to be  problematic   due to  various no-go theorems  (see  \ci{Bekaert:2010hw}). 
 While cubic  higher spin interaction vertices  consistent with 
 on-shell gauge invariance  were  constructed  in various approaches  \cite{Bengtsson:1983pd,
 Berends:1984wp, Metsaev:1991mt, Fotopoulos:2008ka,Boulanger:2008tg,Manvelyan:2010jr,Sagnotti:2010at,Metsaev:2012uy}
 it is not a priori  
 clear if  they can be completed by quartic  and higher   vertices to a {\it local} 
   gauge-invariant  action  that can be used to define  non-trivial  
    observables.
   These issues  were addressed in \cite{Bekaert:2010hp,Taronna:2011kt,Dempster:2012vw}\foot{In particular, ref. 
     \cite{Bekaert:2010hp}  demonstrated the impossibility
     to complete by local quartic vertices  the on-shell gauge-invariant 3-derivative cubic vertex 
     for  three spin-3  fields.}
    and recently in 
   \ci{Ponomarev:2016jqk,talk,Bengtsson:2016hss,Sleight:2016xqq,Ponomarev:2016lrm,tar17}.
 
 Here we shall revisit   the construction  of 
 quartic higher spin interaction vertices 
 for a  minimal  theory   of a single  tower of massless  even  spins $j=0,2,4,...$ 
   (without internal symmetry indices) 
     using the   Lorentz-covariant S-matrix-based   approach. 
   \ar{We shall assume that the theory  should admit a Lorentz-covariant 
   formulation  with  local  on-shell gauge-invariant cubic   vertices 
   and determine the type of non-localities that may appear in the quartic vertices.} 
 We shall  consider the tree amplitudes  involving   three  scalar fields   and one  spin  $j$ field   and show 
 that in the cases  of $j=2$ and $j=4$ 
 there exist  {\it local} quartic $000j$   interaction  terms in the action  that render the  amplitudes  on-shell 
  gauge invariant. However,  for $j \geq 6$ we   shall find that 
  the gauge invariance   requires the introduction of non-local four-point 
  vertices  in the action.\foot{These   conclusions   were reported in \ci{talk}. 
  Similar  results appeared  in  \cite{Taronna:2011kt}  and  also in \ci{tar17}.} 
    
  One may   wonder if the   locality  can  be restored   by  extending the set of fields.     
  \ar{We will suggest that  this may be}  possible    by adding     a second tower of even  spin $j>0$ 
  fields with specific couplings to the fields of the original  set.   
   There are   indications   that   the resulting  extended  interacting action 
   may still  lead,   after the summation over  all intermediate higher spin exchanges,  to    a trivial   S matrix.
 That   would be   in agreement with constraints imposed   by gauge invariance  (under the key assumption  of locality) 
 on  massless    higher spin scattering   amplitudes     that can be  found  in  a  soft momentum limit. 
 
  
  
  A related   question is   about   an  underlying  global  symmetry 
   of  such  conjectured   flat-space massless higher spin theory. While for the 
   higher spin theory in AdS   space   there is a natural higher   spin  symmetry 
    algebra  \ci{Vasiliev:1990bu},  it is    unclear  a priori  if  it has a     flat-space counterpart. 
 By analogy with the Einstein theory that admits a 
  conformal  off shell extension  \ci{frad} 
   (found by introducing a conformally coupled scalar and then solving for it) 
    one may conjecture that there exists a massless  higher spin theory 
   which  is invariant under the  same  conformal higher spin algebra    as  the  conformal higher spin   theory \ci{Fradkin:1985am,Segal:2002gd}. 
   In contrast to the higher-derivative  but local  conformal higher spin    action,  
    the   action   of  the   minimal massless   higher spin 
     theory with two-derivative kinetic terms  for  a single tower of even spins 
     should   be non-local. 
     Its   infinite dimensional   global symmetry     may then  be expected to  constrain  
    the S matrix   to be trivial   like    in the conformal higher spin theory 
     \ci{Joung:2015eny,Beccaria:2016syk}.
  That the scattering amplitudes  of  a massless   higher spin theory  in flat space 
  should vanish   as a consequence of a higher spin symmetry was  also 
  argued for in \ci{Sleight:2016xqq}.

 We shall start in section~2 by discussing the   conditions  that  gauge invariance imposes
 on S-matrix elements, emphasizing that only the 
 ``linear" part of the gauge transformation (i.e. the part independent of the 
  fields) constrains the S matrix.
  We will also 
 review the \ar{Lorentz-covariant local   three-point vertices   we will start with} 
 and   specify the  general form of the quartic Lagrangian
that will be relevant for the calculation of the $000j$ tree-level scattering amplitude.

 In section~3 we  will   compute    the tree-level   $000j$ S-matrix element, 
   finding separately  the exchange part 
 and the contribution of the  quartic  interaction term in the  Lagrangian. We  will then  consider   their 
 gauge transformations and extract the constraints imposed by the  gauge invariance  of the total amplitude 
 on the  coefficient 
 functions   appearing in  the quartic vertex. 

 Section~4 will contain the  analysis    of  
  these constraints. For $j=2$ and $j=4$ we  will 
   find that there exist local quartic terms in the Lagrangian  that  are consistent with the 
  gauge invariance of the  amplitude, while  for $j\ge 6$ a quartic Lagrangian  must be nonlocal.
 %
 
In section~5 we  will   present a  ``minimal"  choice of nonlocal terms required by gauge 
 invariance of the S matrix and  suggest  a way of eliminating the nonlocalities by introducing an
  additional tower of
  ``ghost-like" higher spin  fields. 
  It turns out that the additional
 quartic nonlocal interaction of spin-0 particles required by this  procedure 
  is such that  
 it cancels  the exchange part of  spin-0  four-particle amplitude. 
 Further nonlocal non-minimal terms may completely cancel all the singular terms
 of the exchange part of the  $000j$ amplitude.

 In section~6   we  will  consider  the conformal off-shell  extension of the Einstein theory  
 and show that its perturbative expansion 
 is the same as that of the non-local action resulting from integrating out the trace of the graviton field from the standard Einstein 
 Lagrangian. We  will then conjecture  the  existence of a similar   conformal off-shell extension  of 
 a massless   higher spin theory  that may    have the same symmetries as 
  the  conformal higher spin theory.

 Appendix~A  will contain  some   details of the 
 the contribution of the quartic Lagrangian to the $000j$ amplitude and its organization into a
 basis of contractions of the spin-$j$ polarization tensor used in section 3. 
 
 In Appendix~B we   will   place  the   results of sections  3--5  into a  more  general context
 by   presenting  the analysis  of the constraints  imposed  by the on-shell   gauge invariance 
   and locality   on  generic   massless    higher spin scattering amplitudes $ j_1\dots  j_n j$. 
   Similar  analysis   was performed  earlier   in   \cite{Taronna:2011kt}   with the same conclusions. 
   We shall use the soft  momentum $p_{n+1}\to 0$ expansion 
 generalizing  the   discussions in  \cite{Low:1958sn,Weinberg:1964ew,Bern:2014vva}
  to arbitrary couplings of higher-spin fields. 
 
 In Appendix C   we will  demonstrate that there  exists a special choice of  on-shell gauges
  (or, equivalently, reference vectors 
 in   polarization tensors)  for which the non-local 
 four-point vertex resulting from  integrating out the trace  of the graviton field  in the Einstein action 
  gives a vanishing  contribution   to the  four-graviton amplitude.



\section{Lagrangian vs. S-matrix  gauge symmetries and \\
 massless higher spin interaction vertices }


The construction of  Lagrangians invariant 
 under 
gauge transformations   can be done using  the Noether procedure  but this 
is typically difficult. In this section we   will first  argue that a more efficient and
 technically more straightforward approach 
is to start with  the S matrix  and demand its (on-shell)   gauge invariance. 
We shall then review the   known  \ar{local Lorentz-covariant} cubic interactions of  massless higher-spin fields and write down the most general 
ansatz for   the spin $000j$  quartic Lagrangian which will be used  in the following sections.

\subsection{Gauge transformations from the perspective of the  S matrix}

Lagrangians exhibiting gauge symmetries are usually determined through an iterative Noether procedure.
One starts with a quadratic 
action and deforms it by higher-order terms while simultaneously deforming the  linearized 
gauge transformations in such  a way 
 that the  resulting   action is invariant off-shell under the deformed  transformations. 
This procedure links the construction of the full   action $S= S_2 + S_3 + S_4 + ... $
to the determination  of   a non-linear modification of the gauge transformations  
$ \delta = \delta^{(0)}  +   \delta^{(1)} +  \delta^{(2)}+ ...$. 
 For the cubic part   of the action    one is to 
solve the equation
\be\la{1} 
\delta^{(0)} S_3 + \delta^{(1)} S_2 = 0 \ ,
\ee
where $\delta^{(1)}$ is a deformation of the  
 gauge transformations   linear in the  fields. 
 Thus, the cubic action must be invariant under the  linearized gauge transformations
  up to the term proportional to the 
   free equations of motion. 
   The quartic action $S_4$ is then found  from 
\be\la{2} 
\delta^{(0)} S_4 + \delta^{(1)} S_3 + \delta^{(2)} S_2 = 0 \ .
\ee
Determining  higher $\delta^{(n)}$ 
simultaneously with $S_{n+2}$
is not always straightforward, 
especially in theories with many fields.  

An alternative approach is to constrain the Lagrangian by demanding that the
tree-level scattering amplitudes following from it are invariant 
under the on-shell gauge transformations. 
The essential advantage of this approach is that only 
the 
linearized gauge transformations $\delta^{(0)}$ 
act on   physical scattering amplitudes. While  field-dependent   (``nonlinear")
  terms in the gauge transformation, 
\be
\delta \phi \sim \partial\epsilon +\phi\,  \epsilon +\dots \ ,
\label{gt_schematic}
\ee
relate  $n$-point Green's functions to Green's functions of at least  $
n+1$ fields, such terms are projected out by the amputation 
relating the $n$-point Green's functions and $n$-point scattering amplitudes at generic momenta. 
Indeed, for asymptotic states with momenta
$p_1, \dots,p_n$, the amputation leading to the $n$-point amplitude selects the most singular term
 proportional to $p_1^{-2}\dots p_n^{-2}$ by multiplication with  $p_1^{2}\dots p_n^{2}$   and taking  the on-shell limit $p_i^2 = 0$.
For an  $(n+r)$-point  ($r=1,2,...$) Green's function 
 resulting from  a nonlinear term in the gauge transformations the 
 momentum conservation requires that it   should have a different pole structure. Thus  such 
  terms are amputated away, i.e. 
all the nonlinear terms in the  symmetry   transformations    applied to  Green's  functions are  projected
 out by the LSZ reduction.\footnote{Once a Lagrangian that leads to gauge invariant scattering amplitudes is determined, one 
 may use it to find the nonlinear extensions $\delta^{(1)}, \delta^{(2)},...$ of the linearized gauge transformations. The only 
 changes of the Lagrangian that are still allowed are proportional to the free equations of motion (such terms  can
 be eliminated by field redefinitions   not changing the S-matrix). }

In the context of the  Yang-Mills theory  this  is 
reflected  in that  the amplitudes are invariant under
the  global part of the gauge group and vanish if the polarization vector of a gluon  $\varepsilon_\mu(p)$  is replaced by the momentum
\be 
\delta^{(0)} A_\mu = \partial_\mu\epsilon 
~~ \longrightarrow~~
\delta \varepsilon_\mu(p)  =  p_\mu\, \epsilon(p) \ .
\ee
For   massless 
 higher-spin fields  ($\phi_s \equiv  \phi_{\mu_1\cdots \mu_s} $) the   linearized gauge transformations are given by the first term in \eqref{gt_schematic},  i.e.  symbolically 
\be
\delta^{(0)}  \phi_s = \partial\epsilon_{s-1} \ .
\label{lin_gt}
\ee  
The   corresponding scattering amplitudes for any number of external legs and loop order should  thus be 
invariant under   the    following transformation 
of the polarisation tensor  $\phi_{\mu_1\cdots \mu_s} (p) $    (in  momentum space  representation) 
\be
\delta \phi_{\mu_1\cdots \mu_s} (p) 
                             = p_{(\mu_1} \epsilon_{\mu_2\cdots \mu_s )}(p) \ .
\label{k_gt}
\ee
If the  cubic action $S_3$  is  invariant off shell under  the  linearized gauge transformations \rf{1} 
 then adding a quartic  term $S_4$   is not required by the Noether procedure. 
  In non-trivial   cases  when 
the invariance  of $S_3$ under \eqref{lin_gt}  is only on shell, {\it i.e.} only up to the  free
equations of motion as in  \rf{1}, then  adding $S_4$  is  necessary.
That can be seen   from the S-matrix   perspective  as follows. 
When a three-point vertex is put into  a higher-point amplitude, the inverse propagators generated by its  gauge 
transformation (which vanish on shell) cancel propagators and thus lead to a contact higher-point violation of gauge invariance. 
Repeating the argument implies that, barring special circumstances, vertices of an arbitrarily high  order 
are required.

\subsection{Cubic and quartic terms in a massless higher spin  action }

We shall consider the  totally symmetric  massless  Fronsdal  fields in $d=4$  \cite{Fronsdal:1978rb}
that can be   represented by 
\be\te  \phi_s(x, u)  = \frac{1}{s!}  \phi_{\mu_1\dots\mu_s}(x) \, u^{\mu_1}\dots u^{\mu_s} \ , \la{3} \ee
where $u^\mu$ is an auxiliary  constant vector. 
To construct  cubic  vertices in the covariant form one usually starts by specifying their traceless transverse 
parts. Then these vertices can be promoted to off-shell ones \cite{Manvelyan:2010jr,Sagnotti:2010at,Fotopoulos:2008ka}.
For the calculation of the $000j$ scattering amplitude 
 in the following section (which will  be similar to the one   for $j=0$ in   \cite{Bekaert:2009ud,Ponomarev:2016jqk}) 
 it will be  sufficient to know the vertices in the  de Donder gauge  
\be\te 
\hat D \phi_s(x, u) = 0\ , \qquad 
\qquad
\hat D\equiv 
(\partial_x \cdot \partial_u)-\ha
(u\cdot \partial_x) \partial^2_u 
\label{deDonder}
\ee
To compute   S-matrix elements with  one external higher-spin particle  and 
several   spin 0   ones  
  we need only  cubic  vertices with at  least  one of the fields having  spin 0.  
In this  case   it turns out  that  the traceless-transverse vertices give already the consistent  vertices 
in  the de Donder gauge, i.e. they do not require any completion.
Thus  the   cubic   action  required  for the calculation of the exchange part of the $(000j)$ amplitude is
(see, e.g.,  \cite{Ponomarev:2016jqk} for  details)
\be
S_3[\phi_0,\phi_{j_2},\phi_{j_3}] &=& c_{0j_2j_3}\int d^4 x \Big[(\partial_{u_2} \cdot \partial_{x_{3 1}})^{j_2}
(\partial_{u_3} \cdot \partial_{x_{1 2}})^{j_3}\cr
\label{8}
&&\qquad \qquad \qquad \qquad \qquad 
\times\  \phi_0(x_1)\phi_{j_2}(x_2,u_2)\phi_{j_3}(x_3,u_3)\Big]_{\substack{{u_i=0}\\{x_i=x}}}\ ,
\ee
where
$ \partial_{x_{i j}} \equiv \partial_{x_i}-\partial_{x_j} $.

\ar{Let us note that  here  we will be interested in  determining non-local structures  required by gauge invariance 
 in four-point vertices in a 
 massless  higher spin theory  in flat space   that   has  only manifestly   local and  Lorentz-invariant    cubic 
vertices.  In the direct  light-cone approach developed in  \ci{Metsaev:1991mt} 
 there   are additional lower-derivative cubic couplings,   which are required for the necessary 
  consistency  conditions (Poincar\'e algebra) to be satisfied,  that 
 do not have  local   Lorentz-covariant  counterparts. The light-cone approach  of  \ci{Metsaev:1991mt} 
 need  not  a priori be equivalent   to  an approach based on manifestly covariant local cubic vertices   we are assuming here. 
  It would  still  be interesting 
   to study the role  of these additional  lower-derivative couplings  in the construction of the four-point interaction  Lagrangian 
either  directly in the light-cone approach   \ci{Ponomarev:2016lrm}  or   using their non-local    covariant   versions
  (cf. \ci{Sleight:2016xqq,tar17})  but we expect   that the additional  non-localities associated to them  cannot cancel  against 
  the non-local terms  coming from manifestly covariant cubic vertices \rf{8} 
   that  we shall discuss   below}.\foot{\ar{We   thank R. Metsaev   and D. Ponomarev for  useful discussions
  of  this issue.}}


\ar{Most of  the qualitative   conclusions  below    will not depend on a particular   choice of the coupling constants 
$c_{0j_2j_3}$   in \rf{8}. Still,  to be able to present  closed-form  analytic expressions for the    exchange 
amplitudes (found by summing over all  intermediate spins)   and  thus for the  related   terms in the quartic  vertices 
it is natural to follow 
  \cite{Ponomarev:2016jqk}  and choose $c_{0j_2j_3}$ 
     as in \ci{Metsaev:1991mt}}\foot{\ar{Another  
     motivation  for this choice 
     is that the  same  cubic  coupling  constants 
      appear in the  covariant  higher spin theory  in $AdS_4$   \cite{Sleight:2016dba},  
       suggesting that they  may  also appear  in its  flat-space limit, assuming it  exists.}}
\be
\la{29}
c_{0j_2j_3}=\ g\, \frac{\ell^{j_2+j_3-1}}{(j_2+j_3-1)!}\ .
\ee
Here  $g$  is  an  overall   dimensionless coupling counting the power of fields in interaction vertices 
 and 
$\ell$ is a  unique dimensional  parameter (that will be set to 1  in what follows but can be easily restored on dimensional grounds).
Note that 
$c_{000}=0$, i.e. there is  no cubic scalar  self-coupling.


The on-shell gauge invariance of the cubic vertices implies that the gauge transformation of the 
exchange part of  a four-point 
amplitude  is a local function of momenta.
 It may be cancelled by a  gauge transformation  
of the  contribution   of 
a four-point   vertex   in the Lagrangian, as   it happens  in the case  of the standard  gauge-invariant Lagrangians with spins 
less or equal to 2. 
 Below  we shall explore the possibility of this cancellation  in the case of 
   four-point  $000j$ scattering  amplitudes.

The most general expression for the  $000j$ Lagrangian   written in momentum space    can be represented as follows\foot{We omit the overall momentum  conservation factor $\delta^{(4)}(p_1+p_2+p_3+p_4)$.}
\iffa
\be
\no 
{\cal L}_{000j} &=& \, 
\rC_{j}{}_0(p_1, p_2, p_3)\,  \phi_0(p_1)\,  \phi_0(p_2)\,  (2ip_3\cdot\partial_u)^{j}\phi_0(p_3)\,  \phi_j(p_4, u)
\\
& &+ \rC_{j}{}_{j/2}(p_1, p_2, p_3)\, \phi_0(p_1)\,  
(2ip_2\cdot\partial_u)^{j/2}\phi_0(p_2)\  (2ip_3\cdot\partial_u)^{j/2}\phi_0(p_3)\, \phi_j(p_4, u) \label{21}
\\
& &+
 \sum_{k=1}^{j/2-1}  \rC_{j}{}_k(p_1, p_2, p_3)\,  \phi_0(p_1) \ (2ip_2\cdot\partial_u)^k\phi_0(p_2)\ (2ip_3\cdot\partial_u)^{j-k}\phi_0(p_3) 
\ \phi_j(p_4, u) \ .
\nonumber
\ee
\fi
\be
\label{21}
{\cal L}_{000j} =
 \sum_{k=0}^{j/2}  \rC_{j}{}_k(p_1, p_2, p_3)\,  \phi_0(p_1) \ (2ip_2\cdot\partial_u)^k\phi_0(p_2)\ (2ip_3\cdot\partial_u)^{j-k}\phi_0(p_3) 
\ \phi_j(p_4, u) \ .
\ee
Here $\partial_u$ acts  only on the last factor $\phi_j(p_4, u)$  (cf. \rf{3}) and all $u$-dependence  goes away after the differentiation. 
The  vertex functions $ \rC_{jk}$  ($k=0,1,..., j/2$) are  so far  arbitrary. 
The $k=1$  term  in the sum in \rf{21}   can be set to zero since, up to  a total derivative, it  is equivalent to a shift 
of  $\rC_{j0}$ if $\phi_j$ is taken to be transverse. We will nevertheless keep it for the  symmetry of  the resulting expressions.  
Also,  note that $\rC_{j}{}_0$ is symmetric under the interchange of $p_1$ and $p_2$ 
 while $\rC_{j}{}_{j/2}$ is symmetric 
under the interchange of $p_2$ and $p_3$.  

As was mentioned   above, one may attempt to  determine the quartic Lagrangian
 through the Noether procedure, which links the construction of the Lagrangian to a  nonlinear modification of the gauge transformations.  Instead, below 
  we will   constrain the  coefficient functions $V_{jk}$ in \rf{21}
by demanding   that the S-matrix element $000j$ is  gauge invariant.


\section{The $000j$ scattering amplitude}

In this section we will compute the scattering amplitude of three scalars and one spin $j$ field 
starting  from the Lagrangian 
containing  the standard  Fronsdal kinetic term (in the de Donder gauge) 
 plus the cubic  vertex \rf{8}  and the quartic vertex \rf{21}. 

The momentum (Mandelstam) invariants will be  defined as 
\be
s_{ii'}\equiv (p_i+p_{i'})^2 \ , \qquad 
\qquad
s_{12}+s_{13}+s_{23} = 0 \ .  
\ee
We shall  also use the following notation  for contractions of the spin $j$ field (its Fourier  transform) 
  with  momenta:
\be \la{22}
\phi_j(p, q_1^{k_1},...,  q^{k_n}_n ) \equiv  (q_1\cdot \partial_u)^{k_1} ... (q_n\cdot \partial_u)^{k_n} \phi(p, u)\ , \ \qquad  k_1+...+k_n=j \ .\la{4}   \ee
Thus $\phi_j(p, (c\, q)^j)\equiv (c\, q\cdot \partial_u)^j \phi(p, u)  = c^j \phi_j(p, q^j)$, etc.

\subsection{Exchange contribution and its gauge transformation}

The calculation of the exchange part of the $000j$ S-matrix element from the covariant cubic  action 
\rf{8} follows  \cite{Bekaert:2009ud}  and, especially,  \ci{Ponomarev:2016jqk},
 where  this  was done  for the case of $j=0$  using the action \rf{8}, \rf{29}.
 The  only difference  compared to $j=0$ case comes from the 
presence of the spin-$j$ polarization tensor $\phi_j$  and the coupling of this field. 
The amplitude decomposes in the usual way into  $s$-, $t$- and $u$-channel exchanges,
\be
{\cal A}^{ex} &=& A^{ex}_s+A^{ex}_t+A^{ex}_u \ , \la{32} \\
A^{ex}_s
&=&\frac{2ig^2}{s_{12}} \phi_j(p_4, (2ip_3)^j)\left[{\cal F}_j(y^{(1,2,3,4)}_+)+{\cal F}_j(y^{(1,2,3,4)}_-)\right]\ , 
\label{ex_s}
\\
A^{ex}_t&=&\frac{2ig^2}{s_{23}} \phi_j(p_4, (2ip_1)^j)\left[{\cal F}_j(y^{(2,3, 1,4)}_+)+{\cal F}_j(y^{(2,3,1,4)}_-)\right]\ , 
\label{ex_t}
\\
A^{ex}_u&=&\frac{2ig^2}{s_{31}} \phi_j(p_4, (2ip_2)^j)\left[{\cal F}_j(y^{(3, 1,2,4)}_+)+{\cal F}_j(y^{(3,1,2,4)}_-)\right] \ .
\label{ex_u}
\ee
The functions ${\cal F}_j(y)$ are given by
\be
&&
{\cal F}_j(y) =
\sum_{\ts=\text{0,2,4,...}}\; \frac{\left(-\frac{1}{4} \ell^2 y^2\right)^{\ts}}{(\ts-1)!(\ts+j-1)!} 
= \ha    \left(\ha y \right)^{2-j}\Big(I_j(\ell y)-J_j(\ell y )\Big) \ ,
\label{calF_j}
\ee
where $J_j$ and $I_j$ are the Bessel and the modified Bessel function, respectively, 
and $\ell$ is the scale parameter in \rf{29}  (set to 1  below). 
The arguments  $y_\pm^{(1,2,3,4)}$ of the function ${\cal F}_j$ in the $s$-channel
are  defined  by 
\be
\ha(y_\pm^{(1,2,3,4)}){}^2\equiv 
s_{13} - s_{23} \pm 2\sqrt{ -s_{13}  s_{23}}
=\big(\,\sqrt{s_{13}} \pm \sqrt{s_{12}+s_{13}}\, \big)^2 \ .
\label{y1234}
\ee
The arguments in other channels, $y_\pm^{(3,1,2,4)}$ and $y_\pm^{(2, 3,1,4)}$,  are obtained 
by   relabeling.
It is useful to  introduce the  following notation
\be
 {\cal X}_{123} &\equiv & 2g^2 {{\left[{\cal F}_j(y^{(1,2, 3,4)}_+)+{\cal F}_j(y^{(1,2,3,4)}_-)\right]}}
 = {\cal X}_{213} \ , 
 \cr
 {\cal X}_{231} &=& 2g^2 {{\left[{\cal F}_j(y^{(2,3, 1,4)}_+)+{\cal F}_j(y^{(2,3,1,4)}_-)\right]}}
  = {\cal X}_{321} \ , \la{88}
 \\
 {\cal X}_{312} &=& 2g^2 {{\left[{\cal F}_j(y^{(3, 1,2,4)}_+)+{\cal F}_j(y^{(3,1,2,4)}_-)\right]}}
 = {\cal X}_{132}  \ ,\no 
 ~~~~
\ee
so that \rf{32}   becomes\foot{
For $j=0$ these expressions reproduce the 0000 exchange discussed in \cite{Ponomarev:2016jqk} up to the change of notation $s_{ij} \to  -s_{ij}$.} 
\be
{\cal A}^{ex} = 
\frac{i}{s_{12}} {\cal X}_{123} \ \phi_j(p_4, (2ip_3)^j)  
+\frac{i}{s_{23}} {\cal X}_{231}\ \phi_j(p_4, (2ip_1)^j)
+\frac{i}{s_{13}} {\cal X}_{312}\ \phi_j(p_4, (2ip_2)^j)  \ . 
\ee
Under a gauge transformation  \eqref{k_gt} of the spin-$j$ field   the 
contraction of its polarization tensor with some vector $q$, i.e.  $\phi_j(p_4, q^j)$, becomes
\be
\phi_j(p_4, q^j)\ \ \  \mapsto   \ \ \  j\,  (p_4\cdot q)  \epsilon_{j-1}(p_4, q^{j-1}) \ .
\label{gt}
\ee
In our case  $q$ is the momentum of one of the  spin-0 particles. Because of the momentum 
conservation and transversality of the polarization tensor and the gauge parameter one momentum 
(other than $p_4$) can be eliminated from the contraction with $\epsilon_{j-1}$.
Choosing it to be $p_1$  and using the same notation  as in \rf{4}, i.e. 
denoting by  $\epsilon_{j-1}(p, a^n, b^{j-1-n})$ the contraction of 
$\epsilon_{j-1}$ with a symmetric product of $n$   vectors $a$ and $(j-1-n)$   vectors $b$
we find
\be
&&\delta {\cal A}^{ex}=  
-( {\cal X}_{123} - {\cal X}_{231} ) ~j\, \epsilon_{j-1}(p_4, (2ip_3)^{j-1})
- ({\cal X}_{312} -{\cal X}_{231} )~j\, \epsilon_{j-1}(p_4, (2ip_2)^{j-1})
\cr
&&\quad 
+\, {\cal X}_{231} ~j\Big( ~C_{j-1}^{j/2}  \Big[\epsilon_{j-1}(p_4, (2ip_3)^{j/2-1}, (2ip_2)^{j/2})
+ \epsilon_{j-1}(p_4, (2ip_2)^{j/2-1},(2ip_3)^{j/2})\Big]
\cr
&&\quad\quad\quad
+ \sum_{k=2}^{j/2-1} C_{j-1}^{k-1}\Big[  \epsilon_{j-1}(p_4, (2ip_2)^{k-1},(2ip_3)^{j-k})
+ \epsilon_{j-1}(p_4, (2ip_3)^{k-1}, (2ip_2)^{j-k})\Big]\Big) \ , ~~~~~~~~
\label{gtExchangeFinal}
\ee
where $C_j^k$ are the binomial coefficients (we used that $C_{j-1}^{j-k} =C_{j-1}^{k-1}$ and $C_{j-1}^{j/2-1}=C_{j-1}^{j/2}$).

Let us   note that if the fields were taking  values in the adjoint representation of some internal symmetry 
group, then the $s$, $t$ and $u$ channel contributions   would   be  dressed with 
additional color factors. Gauge invariance would then need
 to be demanded separately for each independent color factor.
  There are
three different partial amplitudes, 
corresponding to the traces  $\Tr[1,2,3,4]$, $\Tr[2,3,1,4]$, $\Tr[3,1,2,4]$, and therefore three 
different gauge transformations that must be cancelled separately:
\be
\delta A_s^{ex}+\delta A_t^{ex}\ , 
\qquad
\delta A_t^{ex}+\delta A_u^{ex}\ , 
\qquad
\delta A_u^{ex}+\delta A_s^{ex} \ .
\ee
In either   abelian or non-abelian case,  the non-trivial   gauge  transformations  of the exchange
 contributions   imply that  a ``contact"  $000j$ term  that should  come 
  from  the four-field    Lagrangian \rf{21} 
is required  to be  added to restore  gauge invariance  of the full amplitude.

\subsection{Contact term contribution and its gauge transformation}

It is straightforward to write down the contribution ${\cal A}^{ct} $ of the  four-vertex  
 \eqref{21} to the $000j$ tree-level amplitude; 
we present  it in Appendix~\ref{contact_term} together with its gauge variation $\delta{\cal A}^{ct} $. Separating the independent contractions of the 
gauge parameter $\epsilon_{j-1}$ and collecting similar  terms we find
\be
\delta{\cal A}^{ct} 
&=& -(j s_{24} {\cal B}_2  + s_{34} {\cal D}_{32,1}  ) \ \epsilon_{j-1}(p_4, (2i p_2)^{j-1})  \no
\\[3pt]
&&  - (j s_{34} {\cal B}_3 +  s_{24} {\cal D}_{23,1}  )\  \epsilon_{j-1}(p_4, (2i p_3)^{j-1})
\nonumber\\[3pt]
&&- \left( \textstyle{\frac{j}{2}} s_{24} {\cal B}_{23} + \left(\textstyle{\frac{j}{2}}+1\right) s_{34} {\cal D}_{23,j/2-1} \right) 
\epsilon_{j-1}(p_4, (2i p_2)^{j/2-1}, (2i p_3)^{j/2})
\cr
&&- \left( \textstyle{\frac{j}{2}} s_{34} {\cal B}_{23} + \left(\textstyle{\frac{j}{2}}+1\right) s_{24} {\cal D}_{32,j/2-1}\right) 
\epsilon_{j-1}(p_4, (2i p_3)^{j/2-1}, (2i p_2)^{j/2})
\label{gtContactFinal}\\
&&- \sum_{k=2}^{j/2-1} 
\big[(j-k+1) s_{34} \,{\cal D}_{23,k-1}+ k  s_{24} \,{\cal D}_{23,k}\big]
\epsilon_{j-1}(p_4, (2i p_2)^{k-1}, (2ip_3)^{j-k})
\cr
&&- \sum_{k=2}^{j/2-1} 
\big[ (j-k+1) s_{24} \, {\cal D}_{32,k-1}+ k s_{34} \,{\cal D}_{32,k}\big]
\epsilon_{j-1}(p_4, (2i p_3)^{k-1}, (2ip_2)^{j-k}) \ .
\nonumber
\ee
The coefficients ${\cal B}$ and ${\cal D}$ are  the combinations of the 
four-vertex    functions $V_{jk}$  defined in eqs.~\eqref{B2}-\eqref{a7} of Appendix~\ref{contact_term}.

Eq.~\eqref{gtContactFinal}  is written under  the assumption  that all the  fields are  singlets (i.e. the theory is abelian).
In the   case   
when   they take 
values in the  adjoint representation of an internal symmetry group 
 the expression in \rf{gtContactFinal}    breaks up into contributions to the three different 
trace structures,  as   in the  exchange contribution discussed above.
This separation may be done by inspecting the explicit coefficients
 given in Appendix~\ref{contact_term} and  assigning the   arguments  of the coefficient 
functions $V_{jk}$ in the four-vertex  \rf{21} as follows:
\be
&& \{(p_1, p_2, p_3), (p_3, p_2, p_1)\} \rightarrow \Tr[1,2,3,4]\ , 
\qquad
\{(p_2, p_3, p_1), (p_1, p_3, p_2)\} \rightarrow \Tr[2,3,1,4]\ , 
\cr
&&\qquad \qquad \qquad \{(p_3, p_1, p_2), (p_2, p_1, p_3)\} \rightarrow \Tr[3,1,2,4] \ .
\ee

\subsection{Constraints  from gauge invariance of the  amplitude}

The gauge invariance of the total  $000j$ amplitude  \be \la{316}
{\cal A} = {\cal A}^{ex}+ {\cal A}^{ct} \ee
demands that  the   variations 
 \eqref{gtExchangeFinal} and \eqref{gtContactFinal} cancel each other, i.e. $\delta {\cal A}= \delta {\cal A}^{ex}+\delta {\cal A}^{ct}=0$.
This leads to the following  constraints on the coefficients ${\cal B} $ and ${\cal D}$ and consequently on the coefficients $V_{jk}$ in the four-vertex term in the Lagrangian \rf{21}.
It  is useful to separate the  $j=2$ and $j=4$ from the general $j$  case.

\noindent
$\bullet$ {$j=2$}:
Here one has  only two structures in the gauge transformation of the $0002$ amplitude: 
$ \epsilon_{1}(p_4,  2ip_2)$  and $ \epsilon_{1}(p_4,  2ip_3)$. The equations following from the vanishing 
of their coefficients in the total variation $\delta {\cal A}$  
 are:
\be
2 s_{13} {\cal B}_2    +  s_{12} {\cal B}_{23} &=& -2{\cal X}_{312} + 2 {\cal X}_{231}\ , 
\nonumber\\[3pt]
2 s_{12} {\cal B}_3   + s_{13} {\cal B}_{23}  &=& -2 {\cal X}_{123} +2 {\cal X}_{231}\ . 
\label{j_eq_2_constraints}
\ee
\noindent
$\bullet$ {$j=4$}: 
Here  there are four independent structures in the  gauge variation of the amplitude: 
$ \epsilon_{3}(p_4,  (2ip_2)^3)$,  $ \epsilon_{3}(p_4,  (2ip_3)^3)$, $ \epsilon_{3}(p_4,  2ip_3, (2ip_2)^2)$ 
and  $ \epsilon_{3}(p_4,  2ip_2, (2ip_3)^2)$. The equations following from the vanishing of their coefficients 
are:
\be
4 s_{24} {\cal B}_2  + s_{34} {\cal D}_{32,1}   &=&  -4 {\cal X}_{312} + 4{\cal X}_{231}\ , 
\nonumber\\[3pt]
4 s_{34} {\cal B}_3  + s_{24} {\cal D}_{23,1}   &=&  -4 {\cal X}_{123} + 4{\cal X}_{231}\ , 
\cr 
2 s_{24} {\cal B}_{23} + 3 s_{34} {\cal D}_{23,1} &=& + 6 {\cal X}_{231}\ , 
\cr    
2 s_{34} {\cal B}_{23} + 3 s_{24} {\cal D}_{32,1} &=& + 6 {\cal X}_{231}\ .
\label{j_eq_4_constraints}
\ee

\noindent
$\bullet$ {$j\ge 6$}: Setting to zero the coefficients in $\delta {\cal A}$  of the terms proportional to 
\be
&&
 \epsilon_{j-1}(p_4, (2ip_2)^{j-1}),\ \epsilon_{j-1}(p_4, (2ip_3)^{j-1}), \ \epsilon_{j-1}(p_4, (2ip_2)^{j/2-1},(2ip_3)^{j/2})\ ,
\\
&&
 \epsilon_{j-1}(p_4, (2ip_3)^{j/2-1}, (2ip_2)^{j/2}), \ \epsilon_{j-1}(p_4, (2ip_2)^{q-1},(2ip_3)^{j-q}), \
 \epsilon_{j-1}(p_4, (2ip_3)^{q-1}, (2ip_2)^{j-q})  ,\ 
\nonumber
\ee
we find:
\be
&&
j s_{24} {\cal B}_2 + s_{34} {\cal D}_{32,1}   = -j ({\cal X}_{312} -{\cal X}_{231})\ , 
\nonumber\\[3pt]
&&
j s_{34} {\cal B}_3 + s_{24} {\cal D}_{23,1}   = -j ({\cal X}_{123}-{\cal X}_{231})\ , 
\cr
&& 
\textstyle{\frac{j}{2}} s_{24} {\cal B}_{23} + \left(\textstyle{\frac{j}{2}}+1\right) s_{34}   {\cal D}_{23,j/2-1}
= + j {\cal X}_{231} C_{j-1}^{j/2}\ , 
\label{GIconstraints}
\\
&& 
\textstyle{\frac{j}{2}} s_{34} {\cal B}_{23} + \left(\textstyle{\frac{j}{2}}+1\right) s_{24} {\cal D}_{32,j/2-1} 
= + j {\cal X}_{231} C_{j-1}^{j/2}\ , 
\cr
&&  
(j-k+1) s_{34} {\cal D}_{23,k-1} + k s_{24} {\cal D}_{23,k}  = + j {\cal X}_{231} C_{j-1}^{k-1} ~\qquad \text{with}~ 2\le k\le j/2-1\ , 
\cr
&&  
(j-k+1) s_{24} {\cal D}_{32,k-1} + k s_{34} {\cal D}_{32,k} = + j {\cal X}_{231} C_{j-1}^{k-1} ~\qquad \text{with}~ 2\le k\le j/2-1\ . 
\nonumber
\ee


\section{Solution of the  gauge invariance constraints \\
on four-vertex  coefficient functions}

In this section we  will  explicitly  solve the  above constraints  for $j=2$ \eqref{j_eq_2_constraints} and $j=4$ \eqref{j_eq_4_constraints} and find the  coefficients $V_{jk}$  in the 
corresponding {  local}  
quartic  terms in the Lagrangian \rf{21} 
that render the $0002$ and $0004$ amplitudes gauge-invariant. 

We will also analyze the general $j \geq 6$ constraints \eqref{GIconstraints} and show that they do not have  solutions corresponding to a {\it local} quartic  $000j$  Lagrangian. 
Relaxing the requirement of locality, in  section  5 we  will  find the leading non-local $000j$ 
interaction  terms  required by gauge invariance  and
discuss their possible   interpretation.

\subsection{$j=2$}

To simplify the discussion let us  use the freedom in the four-field Lagrangian ansatz
 \eqref{21} to set $V_{21}=0$ as the  corresponding term 
 is equivalent, up to a total derivative, to  a shift of $V_{20}$.  Moreover, the left-hand sides of eqs.~\eqref{j_eq_2_constraints} expressed in terms of  $V_{20}$
  can be organized in such  a way that the symmetries of the right-hand sides  become manifest. It then follows that a general solution of \eqref{j_eq_2_constraints} is
\be
V_{20} (p_1, p_2, p_3) &=& -\frac{1}{2 s_{12}} {\cal X}_{123} - \ s_{23}\, 
 \wC_0(p_1, p_2, p_3) \ ,
 \label{genSol2}
\ee
where $\wC_0$ has the properties\footnote{The former is required for the second term \eqref{genSol2} to be a solution of\eqref{j_eq_2_constraints} while the latter is due to the symmetries of $V_{20}$. }
\be\la{42}
\wC_0(p_1, p_2, p_3) = \wC_0(p_3,p_2,p_1)\ , \qquad
~~~~~
\, s_{23}\, \wC_0(p_1, p_2, p_3)=\, s_{13} \, \wC_0(p_2,p_1,p_3) \ .
\ee
It is possible to choose $\wC_0(p_2,p_1,p_3)$ to be such that it cancels  the pole in the first term in eq.~\eqref{genSol2} and 
leads to  a  local Lagrangian. It is therefore natural to express it in terms of the value of ${\cal X}_{123}$ at $s_{12}=0$. Since 
$s_{13}=-s_{23}$ at this point but not away from it, 
different forms  of ${\cal X}_{123}|_{s_{12}=0}$ lead to different expressions for 
$V_{20}$ which differ by local terms. We shall describe two such forms. 
Let us  define the function ${\cal R}_{2}$ 
\iffa 
 given by 
\be
x\, {\cal R}_{2}(x) =  \frac{g^2}{2} \Big(I_2( \sqrt{-8x}) - J_2( \sqrt{-8x})  \Big)\Big|_{x<0} = -\frac{g^2}{2} \Big(I_2( \sqrt{8x}) - J_2( \sqrt{8x})  \Big)\Big|_{x>0}\ ,  \la{43}
\ee
\fi 
which is related to the value of ${\cal X}_{123} $ at  $s_{12}=0$
(cf. \rf{calF_j},\rf{88})
\be 
s_{23}\, {\cal R}_2(s_{23}) \equiv 
{\rm Res}\Big(\frac{ {\cal X}_{123} }{2 s_{12}}, s_{12} = 0\Big) =   -\frac{g^2}{2 } \Big(I_2(\sqrt{-8s_{23}}) - J_2(\sqrt{-8s_{23}})\Big)
   \ .\la{44}
\ee
One possible option is to choose
\be
\rC_{2}{}_0 (p_1, p_2, p_3) &=&
 -\frac{1}{2 s_{12} } {\cal X}_{123} 
 + \frac{1}{2\, s_{12} }  \Big[ s_{13} {\cal R}_{2}(s_{13}) + s_{23} {\cal R}_{2}(s_{23})+  s_{12}{\cal R}_{2}(s_{12})    \Big]  \ ,
\label{final_C0_j_eq_2}
\ee
which leads to the following expression for the total  $0002$ amplitude:
\be
\label{complete_s_eq_2}
{\cal A}= {\cal A}^{ex}  + {\cal A}^{ct}   &=&
 {i}\Big[\frac{\phi_2(p_4, (2ip_3)^2)}{s_{12} }+\frac{\phi_2(p_4, (2ip_2)^2)}{s_{13}}
+\frac{\phi_2(p_4, (2ip_1)^2)}{s_{23}}\Big]
\cr
&&\qquad \times\Big[ s_{13} {\cal R}_{2}(s_{13} ) + s_{23} {\cal R}_{2}(s_{23})+  s_{12} {\cal R}_{2}(s_{12})    \Big]  \ .
\ee
An alternative choice for $\wC_0$  (which also generalizes 
to $j=4$  case) leads to 
  \be
\rC_{2}{}_0 (p_1, p_2, p_3) &=& 
-\frac{1}{2s_{12}} {\cal X}_{123} - s_{23} s_{13}\,  U_2(p_1,p_2,p_3)\ , \la{47}
\\
 U_2(p_1,p_2,p_3) &\equiv & 
    \frac{{\cal R}_{2}(s_{12})}{2s_{13} s_{23}} 
  +  \frac{{\cal R}_{2}(s_{23})}{2s_{12} s_{13}} 
  + \frac{{\cal R}_{2}(s_{13})}{2s_{12} s_{23}} \ .\la{448}
\ee
The corresponding $0002$ gauge-invariant amplitude is then 
\be
{\cal A}&=&+\frac{2i}{3} U_2(p_1, p_2, p_3)  \Big[\phi_2(p_4, 2i (s_{12} p_2 - s_{13} p_3)^2)
					  +\phi_2(p_4, 2i (s_{23} p_3 - s_{12} p_1)^2)
\cr
&&		\qquad\qquad\qquad\qquad			  
					  +\, \phi_2(p_4, 2i (s_{31} p_1 - s_{32} p_2)^2) \Big] \ .
\label{alternate_j2_bose}					  
\ee
One can  check that the poles of this expression match those of eq.~\eqref{complete_s_eq_2}.

It  may  not be surprising that it is possible to find a local 
 quartic $0002$ contact term that renders the $0002$ amplitude  gauge invariant. 
 The analysis in Appendix B  of  gauge  invariance of the  S matrix  using soft 
  limit   does not lead to a non-trivial constraint  on $000j$   amplitude for $j=2$ (and also for $j>2$ 
  as the  $000$ three-point amplitude vanishes automatically, cf. \rf{b99}).  
 Compared to 
 the  Einstein theory coupled to a scalar
   here in addition we have    higher-spin exchange 
 diagrams implying the presence of higher derivative terms in the associated four-point 0002  vertex. 

\subsection{$j=4$}

The interaction of  one  spin-4 and three spin-0 fields is described by the two coefficients in \eqref{21}: $V_{40}$ and $V_{42}$. As in the spin-2 case $V_{41}$ is equivalent, up to a total derivative,   to $V_{40}$.   Let us  note that the structure of the $V_{42}$-dependent part of the Lagrangian \rf{21}, i.e.  
\be
\phi_0(p_1)\ \rC_{42}(p_1,p_2,p_3) \ (2ip_2\cdot \partial_u)^2\phi_0(p_2)\ (2ip_3\cdot \partial_u)^2\phi_0(p_3)\ \phi_4(p_4) \ ,
\ee
implies that only the part of $\rC_{42}$ which is symmetric in $p_2\leftrightarrow p_3$ and antisymmetric in 
$p_1\leftrightarrow p_2$ (and consequently antisymmetric in $p_1\leftrightarrow p_3$) survives. The $p_1\leftrightarrow p_2$ symmetric 
part is a total derivative  that can be ignored.

The solution to eqs.~\eqref{j_eq_4_constraints} is found  by noticing that the first two equations determine ${\cal D}_{23, 1}$ and
${\cal D}_{32, 1}$ in terms of ${\cal X}_{123}$, ${\cal X}_{321}$  in  \rf{88}  and an arbitrary function which is then obtained  from the 
consistency of the last two equations.
 Locality of the Lagrangian also requires that this function exhibits poles whose residue is 
given by the values of ${\cal X}_{123}$ at $s_{12}=0$. Accounting for  all the constraints we find:
\be\la{410}
\rC_{42}(p_1,p_2, p_3) -\rC_{42}(p_3,p_2,p_1) 
=   -g^2(s_{12} - s_{23}) \big[-\te \frac{1}{15} +  2 s_{1 2} s_{13} s_{23}\ U_4(p_1, p_2, p_3) \big]\ , 
\ee
\be 
\rC_{40}(p_1, p_2, p_3)=
%
-\frac{1}{2 s_{12}} {\cal X}_{123} 
-{g^2}   \frac{(s_{12})^2+s_{13} s_{23} - (s_{23}){}^2}{60 s_{12}}
-\frac{1}{2}(s_{13})^2 s_{2 3} ( s_{1 2}  - 2 s_{23}) U_4(p_1,p_2,p_3)\ , 
\no  
\ee
with $U_4(p_1, p_2, p_3)$ given by  (cf. \rf{44}) 
\be
U_4(p_1, p_2, p_3) &=& 
\half  \Big( \frac{1}{s_{13}}  + \frac{1}{s_{23}} \Big) {\widetilde {\cal R}}(s_{12}) + 
      \half   \Big(\frac{1 }{s_{13}}  + \frac{1}{s_{12}} \Big) {\widetilde {\cal R}}(s_{23})+
      \half   \Big(\frac{1}{s_{23}} + \frac{1}{s_{12}} \Big) {\widetilde {\cal R}}(s_{13}) \ , 
\no \\
 {\widetilde {\cal R}}(x)  &\equiv& \te x^{-3} \,  {\cal R}_4(x) - \frac{1}{30}x^{-2}  =O(1) \ 
\label{propRbar}\ , \\
x\,  {\cal R}_4(x)  & \equiv&    -\frac{g^2}{4 x } \Big(I_4(\sqrt{-8x}) - J_4(\sqrt{-8x})\Big) \ .  \la{412}
\ee
The total amplitude,  written  in a manifestly   spin-0 symmetric form,  is then
\be
{\cal A} 
&=&-\frac{2i}{3} U_4(p_1, p_2, p_3)  \Big[\phi_4(p_4, 2i (s_{12} p_2 - s_{13} p_3)^4)
					  +\phi_4(p_4, 2i (s_{23} p_3 - s_{12} p_1)^4)
\cr
& &		\qquad\qquad\qquad	\qquad 		  
					  +\ \phi_4(p_4, 2i (s_{31} p_1 - s_{32} p_2)^4) \Big]
\cr 
& &+  \frac{i}{45}\Big[ \frac{(s_{12}){}^4+ (s_{13}){}^4}{s_{12} s_{13} s_{23} }  \phi_4(p_4, (2i p_1)^4)
    + \frac{(s_{12}){}^4+ (s_{23}){}^4}{s_{12} s_{13} s_{23} }  \phi_4(p_4, (2i p_2)^4)
\cr
& &		\qquad\qquad\qquad	
   +  \frac{(s_{13}){}^4+ (s_{23}){}^4}{s_{12} s_{13} s_{23}}  \phi_4(p_4, (2i p_3)^4)\Big]
\cr
& & - \frac{2i}{15}\Big[  \frac{s_{13} s_{23}}{s_{12}} \phi_4(p_4, (2i p_1)^2, (2i p_2)^2) 
   +    \frac{s_{12}  s_{23}}{s_{13}} \phi_4(p_4, (2i p_1)^2, (2i p_3)^2) 
\cr
& &		\qquad\qquad
   +  \frac{s_{12}\, s_{13}}{s_{23}} \phi_4(p_4, (2i p_2)^2, (2i p_3)^2) \Big]
\cr
& &+ \frac{4}{45}\Big[ \frac{(s_{23}){}^2}{s_{12}} \phi_4(p_4, 2i p_1, (2i p_2){}^3)
 + \frac{(s_{13}){}^2}{s_{12}} \phi_4(p_4, (2i p_1){}^3, 2i p_2) 
\cr
& &\qquad \quad   + \frac{(s_{23}){}^2}{s_{13}} \phi_4(p_4, 2i p_1, (2i p_3)^3)
 + \frac{(s_{12}){}^2}{s_{13}} \phi_4(p_4, (2i p_1){}^3, 2i p_3) 
\cr
& & \qquad\quad  + \frac{(s_{13}){}^2}{s_{23}} \phi_4(p_4, 2i p_2, (2i p_2)^3)
  + \frac{(s_{12}){}^2}{s_{23}} \phi_4(p_4, (2i p_2){}^3, 2i p_2) \Big] \ .
\la{4122}
\ee
%
While this  expression  superficially contains products of multiple $s_{ik}$ denominators, momentum conservation
implies that not only  all of its poles   are at physical values (i.e. at vanishing Mandelstam invariants) 
but also  the corresponding residues are  local.
%

\subsection{$j\ge 6$}

The analysis of the $j \ge 6$  gauge invariance constraints  \eqref{GIconstraints}
 can be done  in  three steps: 
  solve the  first two equations
for ${\cal D}_{23,1}$ and ${\cal D}_{32,1}$ in terms of a free function, as for $j=4$;  then  solve the last two recursion relations;  
  finally,  use the consistency of the third and the fourth equation to determine the remaining function.

The general solution of the first two equations  in \eqref{GIconstraints} reads:
\be\la{414} 
{\cal D}_{23,1} =-\frac{j}{s_{23}} {\cal X}_{231} -\ s_{12}\, \wC(p_3,p_2,p_1)
\ , \qquad \quad 
{\cal D}_{32,1} =-\frac{j}{s_{23}} {\cal X}_{231} -\ s_{13}\,\wC(p_2,p_3,p_1)\ , 
\ee
where the symmetries of the equations require that
\be
\wC(p_1,p_2,p_3) = \wC(p_3,p_2,p_1) \ .
\ee
This function must be chosen to cancel the pole in the first term in \rf{414}; as in the $j=2$ and $j=4$ cases, we choose its 
pole part to be proportional to the residue ${\cal R}_j$ of the first term (cf. \rf{44},\rf{412}):
\be
\label{Ctilde}
 \wC(p_1,p_2,p_3) =2 j  \big( \frac{1}{s_{12}} + \frac{1}{s_{23}}    \big) \,  {\cal R}_j(s_{13})   +\yg(p_1,p_2,p_3)+\yg(p_3,p_2,p_1)\ , 
\\
{\cal R}_j(s_{23}) \, s_{23}  \equiv  {\rm Res}(\frac{{\cal X}_{123}}{2 s_{12}} , s_{12} = 0) 
=  - \frac{g^2}{2 (2s_{23})^{j/2-1}} \Big[I_j(\sqrt{-8s_{23}}) - J_j(\sqrt{-8s_{23}})\Big]\ .\la{416}
\ee
The   function $\yg$  should  be chosen to be  such that the remaining equations 
are also solved.

The solution of the last two recursion relations in \eqref{GIconstraints}  
is unique:
\be
{\cal D}_{23,k} &=& \frac{(-1)^{k-1}}{j} C_j^k \Big(\frac{s_{12}}{s_{13}}\Big)^{k-1}{\cal D}_{23,1} 
+\frac{1}{s_{13}}\, C_j^k\; {\cal X}_{231}\; \sum_{n=0}^{k-2}  \Big(-\frac{s_{12}}{s_{13}}\Big)^{n}\ , 
\\
{\cal D}_{32,k} &=& \frac{(-1)^{k-1}}{j} C_j^k \Big(\frac{s_{13}}{s_{12}}\Big)^{k-1}{\cal D}_{32,1} 
+\frac{1}{s_{12}}\, C_j^k\; {\cal X}_{231}\; \sum_{n=0}^{k-2}  \Big(-\frac{s_{13}}{s_{12}}\Big)^{n} \ ,
\label{Sol_rec_rel}
\ee
where,  as in  \eqref{GIconstraints},    $C_j^k$  are  the binomial coefficients.

As  the last step, the third and fourth equations in \eqref{GIconstraints} both determine the remaining coefficient ${\cal B}_{23}$;  
demanding that the two solutions are the same, i.e. 
\be
 \frac{(s_{12})^j}{(s_{13})^{j/2-2}} \Big[ (1-(-1)^{j/2})  \frac{1}{s_{23}} {\cal X}_{231}
+ 2\frac{s_{13}}{s_{23}}{\cal R}_j(s_{13})-\frac{s_{12}}{j}\big(\yg(p_1,p_2,p_3)+  \yg(p_3,p_2,p_1)\big)\Big] 
\cr
=
\frac{(s_{13})^j}{(s_{12})^{j/2-2}} \Big[(1-(-1)^{j/2})  \frac{1}{s_{23}} {\cal X}_{231}
+ 2\frac{s_{12}}{s_{23}}{\cal R}_j(s_{12})-\frac{s_{13}}{j}\big(\yg(p_2,p_3,p_1)+\yg(p_1,p_3,p_2)\big)
\Big]\ \ 
\label{420}
\ee
should fix  the remaining  function $Y$ in eq.~\eqref{Ctilde}. 

It turns out that there is  no  {\it local}  (i.e. containing only positive powers of momenta) 
  solution for the function $Y$  and thus for the coefficient  functions $V_{jk}$ in the 
  four-vertex \rf{21}  in   the  higher spin action.
This is  essentially  due to  a  too high power of 
the  Mandelstam invariants which needs to be compensated for the two sides of the equation \rf{420}  to be equal.
To see this explicitly it is sufficient to  consider  the case of $j=6$  when we get 
\be\te 
{\cal R}_6(x) =  \alpha  g^2 x + {\cal O}(x^2) \ ,  \quad 
\quad
{\cal X}_{123} =\frac{1}{4}\alpha g^2 (s_{12}^2 - 8 s_{13}s_{23})+{\cal O}(s_{ij}^4) \ , \qquad  \alpha = \frac{1}{1260}\ ,
 \ee
so that  eq.~\eqref{420} becomes
\be
&&\ \te  3 \alpha  (s_{12}-s_{13})\big[ (s_{12}-s_{13})^4 -2 s_{12}s_{13}(s_{12}^2+s_{13}^2)\big]
\no \\
&& =s_{12}^5 \big[\yg(p_1,p_2,p_3)+  \yg(p_3,p_2,p_1)\big] - s_{13}^5 \big[\yg(p_2,p_3,p_1)+\yg(p_1,p_3,p_2)\big] \ .
\ee
Since  the left-hand side contains terms with powers of $s_{12}$ and $s_{13}$ smaller than 5, $\yg$ must  contain
$s_{12}^{-1}$ and $s_{13}^{-1}$  factors. As $\yg$  enters (through \rf{Ctilde})  the expression for ${\cal D}_{23,1} $ 
in \rf{414}  and thus, via \rf{a6}, the $V_{jk}$    functions in \rf{21}, 
the $000j$ Lagrangian \rf{21} cannot be local for $j\ge 6$.

\section{Non-local  terms in $000j$ vertex  for   $j \ge 6$ \label{linearize}}

The above analysis  implies that 
for $j\ge 6$ there is no local quartic Lagrangian that renders the  $000j$ amplitude gauge invariant.
Let us now   discuss  in detail the  structure of the required 
non-local    terms  in  the corresponding four-point interaction vertex
and attempt to suggest  their  possible  interpretation. 

 Rather than finding the  complete 
solution of the system \eqref{GIconstraints},  it is more convenient to  first make an  
 ansatz for the non-local part of the  functions  $\rC_{jk}$ in  \rf{21} and  then 
determine the  numerical coefficients of various possible  terms  (ignoring  all    local  contributions).

There are, in fact, many nonlocal solutions to the gauge invariance constraint equations.
 It turns out that it is possible to choose 
all  the coefficient functions $\rC_{jk}(p_1,p_2,p_3)$  in \rf{21} with $k >0$  to be local. Moreover, it is possible to choose 
$\rC_{j0}$ to have only a finite number of non-local terms.
 Below we list  representative  expressions  for  $  g^{-2} \rC_{j0}(p_1,p_2,p_3) $
for   $j=6,\dots, 12$:
\be
\begin{array}{r | c}
j &  - g^{-2} \rC_{j0}(p_1,p_2,p_3)   \cr
\hline
 6 &  \frac{1}{2^2 \times 3! \times 7!!}\frac{s_{13}^2+s_{23}^2}{s_{12}} + \text{local} \cr
 8 &  \frac{1}{2^3 \times 4! \times 9!!}\frac{s_{13}^2+s_{23}^2}{s_{12}} 
     + \frac{1}{2^2 \times 3!\times 5! \times 11!!}\frac{s_{13}^4+s_{23}^4}{s_{12}}  + \text{local}  \cr
10&  \frac{1}{2^4 \times 5! \times 11!!}\frac{s_{13}^2+s_{23}^2}{s_{12}}  
     + \frac{1}{2^3 \times 3! \times 6! \times 13!!}\frac{s_{13}^4+s_{23}^4}{s_{12}}
     + \frac{1}{2^2 \times 5!\times 7! \times 15!!}\frac{s_{13}^6+s_{23}^6}{s_{12}}  + \text{local}  \cr
12&  \frac{1}{2^5 \times 6! \times 13!!}\frac{s_{13}^2+s_{23}^2}{s_{12}}      
     + \frac{1}{2^4 \times 3! \times 7! \times 15!!}\frac{s_{13}^4+s_{23}^4}{s_{12}} 
     + \frac{1}{2^3 \times 5!\times 8! \times 17!!}\frac{s_{13}^6+s_{23}^6}{s_{12}} 
     + \frac{1}{2^2 \times 7!\times 9! \times 19!!}\frac{s_{13}^8+s_{23}^8}{s_{12}}  + \text{local}  \cr
\end{array}
\label{examples}
\ee
For generic $j\ge 6 $     the  corresponding choice of the non-local part of the four-vertex   function 
 is  
\be
\rC^{\text{nonloc}}_{j0}(p_1,p_2,p_3) =
 - \frac{g^2}{s_{12}}  \sum_{l=0}^{(j-6)/2} 
\k_{jl}  \big(s_{13}^{2l+2} + s_{23}^{2l+2}\big) \ , \ \ \ \ \ \ \la{52} \\
\k_{jl} = 
\frac{1}{2^{j/2-1} l! (2l+1)!!(l+{j \ov 2})!(2l+j+1)!!} \ .\la{53}
\label{nonlocalVj0}
\ee
%
Transforming to position space, the nonlocal part of the $000j$ Lagrangian \rf{21} is  then 
\be
{\cal L}_{000j}^{\text{nonloc}}= 
 g^2  \sum_{l=0}^{(j-6)/2}
2^{2l+2} \k_{jl}\ 
\phi_0\,  (\partial^{\mu_1}\dots \partial^{\mu_{2l+2}}\phi_0)\  \frac{1}{\Box}\ 
\big[\partial_{\mu_1}\dots \partial_{\mu_{2l+2}}(\partial_u\cdot\partial)^{j}\phi_0\big]\phi_j( u) \ , 
\label{54}
\ee 
where  all fields  have $x$-space arguments and $ \phi_j(u) \equiv \phi_j(x, u)$.

One may  notice  that the numerical coefficient of each term in the sum \rf{54}   factorizes  as
\be
g^2 2^{2l+2} \k_{jl}    =  \cC_{jl}\ \cC_{0l} \ ,  \qquad
\cC_{jl} \equiv  \frac{\sqrt{8}  g}{2^{{j/2}-l}(l+{j \ov 2})! (2l+j+1)!!} \ .  \la{55}
\ee
Then summing  \rf{54} over all $j=6,8,10,...$  we find (after  a  change of    summation  index) 
\be
\sum_{j=6,8,...} 
  {\cal L}_{000j}^{\text{nonloc}} = 
   \sum_{l=0}^\infty \cC_{0l}\, \phi_0\,  (\partial^{\mu_1}\dots \partial^{\mu_{2l+2}}\phi_0 )
\frac{1}{\Box}
 \sum_{j=6+2l}^{\infty} \cC_{jl}  \big[\partial_{\mu_1}\dots \partial_{\mu_{2l+2}}(\partial_u\cdot\partial)^{j}\phi_0\big]\phi_j(u) \ . 
\ \ \label{56}
\ee 
This remarkable factorization 
suggests that it may be  possible  to eliminate the non-locality in the four-vertex  by introducing an additional 
family of spin\footnote{
One may understand the absence of a spin-0 field in this family based on the momentum dependence of the three-field couplings.
Assuming that the 
triple scalar coupling is nonvanishing and constant, a scalar exchange in, e.g., the $s$ channel of a $000j$ amplitude 
contributes $\phi_j(p_4, (2ip_3)^j)/s_{12}$ and its gauge transformation is just $\epsilon_{j-1}(p_4, (2ip_3)^{j-1})$. 
However, since there is no triple-scalar interaction among the original  minimal set of fields, the
 minimal-field exchange starts out as 
$\phi_j(p_4, (2ip_3)^j) P_2(s_{12}, s_{23})/s_{12}$, where $P_2$ is a polynomial of degree 2.  Therefore, there is 
no term in the gauge transformation of the minimal-field exchange part that can be cancelled by a quartic term which 
is equivalent to a scalar field exchange.
} 
$j=2l+2 =2,4,6, ...$ fields  $\psi_{j}$ 
each coupled  to the original $\phi_j$  fields with $j > 2l+6$ \  ($l=0,1,2,...$).
Since for fixed spin $j$ the Lagrangian ${\cal L}_{000j}^{\text{nonloc}}$ has a finite number of singular terms, only 
a finite number ($j-4$, cf. eq. \eqref{54}) of these additional fields contribute to it.

The local  hermitian Lagrangian that reproduces the nonlocal terms in  \eqref{56} 
upon integrating out  $\psi_j(u) \equiv \psi_j(x,u)$   fields may  be written symbolically as
\be
&&{\cal L}_\text{extra} (\phi, \psi) = - \ha  \sum_{l=0}^{\infty} \psi_{2l+2} \Box \psi_{2l+2} \no 
\\
&& - \sum_{l=0}^{\infty} \Big[ \cC_{0l}\, \phi_0 (\partial\cdot\partial_v)^{2l+2}\phi_0  
+\sum^\infty_{j= 2l+6} \cC_{jl}\, \big((\partial_u\cdot\partial)^{j} (\partial_v\cdot\partial)^{2l+2}\phi_0\big) \phi_j(u)\Big]\psi_{2l+2}(v) \  .
  \la{577}
\ee
Note that the additional fields $\psi_j$  are ghost-like 
 -- they  have an unphysical sign of their kinetic terms.

\ar{Assuming    that a consistent action  given  by \rf{577}   together with further local terms depending on $\phi_j$ and $\psi_j$   indeed 
exists   and  defines a local 
 higher spin theory,   then  integrating out $\psi_j$  from \rf{577}   one finds  also 
  other  four-point nonlocal terms (in addition to \rf{56}).}
   These    do not contribute to the $000j$ amplitudes for  $j\ge 0$
and thus are not constrained  by  our  previous analysis.  Explicitly, they are 
\be
&&{\cal L}^{\text{nonloc}}_{0000} = \ha \sum_{l=0}^\infty   (\cC_{0l} )^2\ 
\phi_0 \partial^{\mu_1}\dots \partial^{\mu_{2l+2}}\phi_0\  \frac{1}{\Box}\ \phi_0 \partial_{\mu_1}\dots \partial_{\mu_{2l+2}}\phi_0
\ ,\la{58}  \\
&&
{\cal L}^{\text{nonloc}}_{00(2\times\text{hs})} =  \ha \sum_{l=0}^\infty 
\sum^\infty_{j_1=  2l+6}\sum^\infty_{j_2= 2l+6}
\cC_{j_1l}  \ \cC_{j_2 l} 
 \big[ (\partial_{u_1}\cdot\partial)^{j_1}\partial_{\mu_1}\dots \partial_{\mu_{2l+2}}\phi_0\big] \phi_{j_1}(u_1)
 \no 
\\
& &\qquad \quad \qquad \quad\qquad \quad\times 
\frac{1}{\Box}
\big[(\partial_{u_2}\cdot\partial)^{j_2}\partial_{\mu_1}\dots \partial_{\mu_{2l+2}}\phi_0\big] \phi_{j_2}(u_2)\ ,
\label{59}
\ee
where ${\cal L}^{\text{nonloc}}_{00(2\times\text{hs})}$ are further quartic nonlocal interactions of two spin-0 fields with any 
two fields of higher spin. 
Thus the validity of \rf{577}   rests on the conjecture 
 that these non-local quartic terms are  indeed  present in the minimal higher spin theory
  and  have precisely the right coefficients to have the interpretation of 
  exchange contributions  of the second tower of
   ghost-like fields $\psi_j$. 
 While the four-scalar interaction \rf{58}
cannot be  found   from   the requirement of  gauge invariance, 
the study of the gauge invariance of the  $00j_1j_2$ amplitude   may,  in principle,    confirm  the presence of  \rf{59}.

If   the non-local   four-scalar  vertex  \rf{58}  is indeed present in the minimal higher spin theory,  
we   may  find  its contribution to the  $0000$  amplitude discussed   in \ci{Ponomarev:2016jqk}. 
  Since 
\rf{58}   represents  only the  non-local (pole) 
 part of the quartic  vertex  we  are able to  determine   only  the pole part  of the full four-scalar  amplitude (in each channel). 
Using   that at $s_{12}=0$  one has   $y^{(1,2,3,4)}_-=0$ and $y^{(1,2,3,4)}_+=\sqrt{8 s_{13}}$ (see \rf{y1234}) 
the  pole part of the $s$-channel exchange 
 (see \rf{ex_s},\rf{calF_j})
  and contact contributions  are,  respectively,  
\be
(A^{ex}_s)_{0000}\Big|_\text{pole}&=&
-\frac{2ig^2}{s_{12}} ~ s_{13}\big[  I_0(\sqrt{8 s_{13} }) - J_0(\sqrt{8 s_{13}})\big] \ , \la{510}
\\
(A^{ct}_s)_{0000} \Big|_\text{pole} &=& 4i  \sum_{l=0}^\infty (\cC_{0l})^2 (\ha s_{13})^{2l+2} =  2i g^2 \frac{s_{13}}{s_{12}}
\big[I_0(\sqrt{8s_{13}}) - J_0(\sqrt{8s_{13}})\big]\ . \la{511}
\ee
We observe that the sum of eqs.~\rf{510} and \rf{511} vanishes, i.e. 
 the nonlocal $0000$  vertex whose presence is required   to eliminate the 
 non-local $000j$ vertex    by introducing an extra tower
 of fields $\psi_j$ 
leads to the cancellation of the $s$-channel pole in  the $0000$ amplitude.
By  relabeling of momenta 
one  then  finds that  all  the  poles in other channels of four-scalar  amplitude 
cancel out.
  It is then natural to conjecture that the full  tree-level   $0000$ amplitude   is actually vanishing
  (cf. Appendix \ref{B}).
 
Superficially,   it may seem   that a similar cancellation can not occur for the $000j$ amplitude with $j\ge 6$: the  residue of the 
nonlocal part of $V_{j0}$ in \rf{52} is a polynomial  while the residue of the $s$-channel exchange \rf{ex_s} is a more complicated 
function. Closer inspection shows that the first $(j-4)/2$ terms in the expansion of the exchange part exactly cancel against the 
nonlocal part of $V_{j0}$. Moreover, as noted above eq.~\eqref{examples}, 
the gauge invariance allows further infinitely many nonlocal terms 
in $V_{j0}$; in writing eq. \eqref{examples}  such ``non-minimal" terms  were  set to zero.
 It is in principle possible to make other
 choices, in particular,   such that the complete pole part of the $000j$ S-matrix element cancels out.

It is interesting to consider from this perspective the cases of $j=2$ and $j=4$ for which  we  have 
found a local quartic
Lagrangian rendering the corresponding $000j$ amplitudes gauge invariant. 
In these cases too the solution to the gauge invariance 
constraints was not unique; it is possible that  one can  add   some    ``non-minimal" singular terms 
 that   cancel  the  poles of the exchange part of the 
amplitude.  An indication that this  may be  possible is that, 
up to local terms, the pole part is gauge-invariant.\foot{One can see this by choosing 
a momentum configuration in which one Mandelstam invariant is close to zero; for such momenta the amplitude is dominated 
only by its pole part 
which, consequently, is to be gauge invariant.} Thus, by adding non-minimal nonlocal terms to the quartic 0002 
and 0004 Lagrangian it should be possible to set to zero the pole part of the corresponding amplitudes.


\ar{Under the conjecture    that a local and gauge-invariant  (but non-unitary)  action for  an
 extended set of  higher spin fields  may indeed be constructed, 
the  above  observations may be considered as a hint that  such a  
 theory  
 may  have a  trivial 
S matrix;   
 this    would    be   consistent with 
   expectations  based on  no-go theorems (cf.  Appendix B).  }

\def \na {\nabla}
\def \lam {\lambda} 
\def \J {{\cal J}} 
\def \d {\partial} 
\def \h {{\rm h}}
\def \tmn   {$t_{mn}\ $ }
\def \te {\textstyle} 
\def \foot {\footnote}
\def \four{{\textstyle {1\ov 4}}}
 \def \third { \textstyle {1\ov 3}}
\def\det{\hbox{det}}
\def \fo {\four}
\def \z {\zeta}
\def \p  {\phi}
\def \del {\d}

\section{Non-local actions  and  conformal off  shell extension: \\
spin 2  example
}

To explore    types  of non-localities  in four-point vertices 
that may appear  in  gauge  theory actions   leading to 
 consistent S-matrices   and   also to try to uncover  possible   higher symmetries associated with such actions, 
 it is useful to discuss   first   
 the spin 2   example  of  the Einstein's theory  as   it  may suggest  possible 
 higher   spin generalizations.
 
 The Einstein theory   describes  a massless spin 2  particle  
  by a
reducible Lorenz representation --  symmetric tensor $h_{mn}= g_{mn} - \eta_{mn}$.\foot{In this section $m,n,...$ will 
stand for 4d Lorentz indices. We will also assume  summation over repeated indices   regardless  of their position.} 
Expanded near flat space its  local   action $S_E(h)= \int d^4x \sqrt g R$  
  depends on  both  traceless  $t_{mn}$ and  trace $\h$   parts  of $h_{mn}$
  \be  \la{61} 
  h_{mn} = t_{mn}  + \fo \eta_{mn} \h \ , \ \ \ \   \ \ \ \ \ \ \    t_{mn} \equiv  h_{mn} - \fo \eta_{mn} \h \ , \ \ \ \ \   \h \equiv  h^m_m \ . 
  \ee 
However,  $\h$   may be viewed  as   unphysical --  it   can be gauged away on shell  and thus does  not appear 
as  an asymptotic state in the S matrix.\foot{To recall,  the linearized Einstein equations
   imply  that  $\d^2  h_{mn} =0, \   \d^2  \h=0$  with the 
   gauge  condition $\d_m h_{mn} =0$  assumed. 
   The residual on-shell gauge transformations $\delta  h_{mn} = \d_m \xi_n + \d_n \xi_m$   subject to $\d^2 \xi_n=0, \ \ \d^2  \d_m \xi_m =0$  allow setting $\h=0$  and removing
    all but the  two  remaining degrees of freedom 
   from $t_{mn}$. 
   Thus one can gauge fix $h_{mn}$  to be both transverse and traceless  by  an ``on-shell gauge", 
   but this is not possible off shell. 
   } 
   Splitting $h_{mn}$ into $t_{mn}$ and $\h$   and observing that 
    only  $t_{mn}$ (subject to $\d^2 t_{mn}=0, \ \d_m t_{mn}=0$)
    may appear  on external lines in  the S  matrix
     one may  first  integrate out $\h$. 
     This gives     an effective  action for $t_{mn}$   which 
      is    {\it non-local}
     $\bar S_E(t) = \int d^4x (  t \d^2 t + \d \d t  \d^{-2} \d \d   t +  \d^2 t t t   +  \d^2 t \d^{-2} \d t \d t    + \d^2 t t t t  + \d t \d t  \d^{-2} \d t \d t + ... )$  
     and which   should  produce 
   the same S matrix for gravitons as  in the Einstein's theory. In the transverse gauge $\d_m t_{mn}=0$ 
  the  non-localities in   $\bar S_E(t)$ will start  only at  $t^4$ order.    
   
This action  can be written in a   closed form $\int d^4 x \sqrt g \big(  \te R - 
{1\ov 6}  R \Delta^{-1}   R \big) $  as it
 is   related  to  the 
    Weyl-invariant off   shell  extension of the Einstein's
     theory \ci{frad}. To find it one may  start with  a conformally coupled 
     scalar  action, fix  the Weyl symmetry by choosing the gauge $\h=0$  and then  solve for  the ``compensator" scalar field. 
There is an    analogy  with  the Weyl  $C^2$ gravity  where  the
 Weyl symmetry  is present  off shell  so that  
expanding near flat space  and fixing this  symmetry  by the $\h=0$ gauge  one  also gets   an action 
for $t_{mn}$ only, which  here is 
local  but higher-derivative one: \  $S_W(t) = \int d^4 x \sqrt g\ C^2 = \int d^4 x ( t \d^4 t   + \del^4 ttt + \del^4 tttt + ...) $.

\ar{One may then attempt to   generalize the     above  construction   to the higher spin case:} starting with a  Lorentz-covariant 
 quadratic plus cubic action for the  Fronsdal higher spin totally symmetric  fields $\p_s \equiv (\p_{m_1 ... m_s})$ 
 which are subject to the standard double-traceless condition one  may   
   split them into  the  ``physical" traceless    $t_{s}$   and the ``ghost-like" 
 trace $\h_{s-2} $ parts  and then  integrate  out $\h_{s-2}$. The resulting  non-local 
 action for $t_s$   should then  lead to the same  S matrix   as the original   action for the Fronsdal fields 
 and  may  be   related  to    a higher spin analog of   
   the conformal extension of the Einstein theory. 
   The corresponding higher derivative  counterpart 
  will be the conformal higher spin theory  invariant under the infinite dimensional 
  conformal higher spin symmetry  \ci{Fradkin:1985am,Segal:2002gd}.

\subsection{Integrating out the trace   from  the Einstein action }

Let   us start by recalling  the  near-flat-space expansion of the    Einstein Lagrangian   with $h_{mn}$ split as  in \rf{61}\foot{The expansion of  the  Einstein action to quartic order in $h_{mn}$  appeared, e.g.,  
in \ci{dewit,beren,gorof}  and   the computation of  the four-graviton tree-level S matrix was discussed  in \ci{san}.}  
\be \la{a1}
&&\te  L_E(h)= \sqrt{ g} R = X_1 + X_2 + X_3 + X_4 +  ...\ , \qquad   \qquad  X_n = O(h^n)   \ , 
\qquad \quad \\
&& \te  X_1= \d_m \d_n  h_{mn}  - \d^2 \h  = \d_m \d_n  t_{mn}  - {3 \ov 4} \d^2 \h \la{a2} \ ,  \\
&&
\te    X_2 =  {3 \ov 4} \d_k t_{mn} \d_k t_{mn}  - \ha \d_k t_{mn} \d_n t_{mk}   + t_{mn} \del^2 t_{mn}
- \d_n t_{kn} \d_m t_{km}  -2 t_{mk} \d_k \d_n t_{mn} \no \\
&&\te
\qquad + {3 \ov 32} (\d_k \h)^2  + \fo \d_m t_{mn} \d_n \h +  \ha   t_{mn} \d_m \d_n \h   \la{a3} \ , \\
&& \te 
X_3=-\big(  -  {3 \ov 4} t_{mn} \d_m t_{sr} \d_n t_{sr}   +   t_{ms} \d_m t_{nr} \d_n t_{sr} 
+ \ha  t_{ns} \d_m t_{nr} \d_r t_{sm} - {3\ov 2}  t_{ns} \d_m t_{nr} \d_m t_{sr} 
\no \\
&& \te \qquad
+  t_{mr}  t_{ns}  \d_m \d_n t_{sr} -   t_{mn}  t_{sr}  \d_m \d_n t_{sr}
-  t_{mr}  t_{ms}  \d^2 t_{rs} \no \\
&& \te \qquad
- \fo t_{rk} t_{rk} \d_m \d_n t_{mn}  +2 t_{mn} \d_m t_{nr} \d_s t_{sr} 
+2 t_{mn}  t_{ms}\d_n \d_r t_{sr}    +   t_{ns} \d_m t_{mn} \d_r t_{sr} \no \\
&&
\qquad \te
-{1\ov 16} t_{mn} t_{mn} \d^2 \h  - \fo t_{mn} \d^2 t_{mn} \, \h
+\fo t_{mr} \d_n t_{nr} \d_m \h  - \ha t_{ nr} \d_m t_{nr} \d_m \h - {3 \ov 16}( \d_m t_{nr})^2 \h\no
 \\ && \qquad \te 
+ {1\ov 8} \d_n t_{mr} \d_m t_{nr} + \fo  \d_m t_{mn} \d_r t_{rn} \h 
+ \ha t_{mr} \d_m \d_n t_{nr} - \fo t_{mn} t_{mr} \d_n \d_r \h \no \\ &&\qquad \te - {5 \ov 32} t_{mn} \d_m \h \d_n \h - {1\ov 16} \d_m t_{mn} \h \d_n \h    - {1\ov 8} t_{mn} \h \d_m\d_n \h- {3\ov 128} \h \d_m \h \d_m \h \big) \la{aa4}\ . 
\ee 
Solving for $\h$ at the classical level will then give  a  Lagrangian depending only on $t_{mn}$: 
\be  
\la{aa3} 
\bar L_{E} (t) = \bar  L^{(2)}_{E} (t) + \bar  L^{(3)}_{E} (t)  +  L^{(4)}_{E} (t)   + ... \ . 
\ee
Explicitly,  the   quadratic part of the    Lagrangian \rf{a1}   is 
(dropping  total derivatives in $X_2$)  
\be \te \la{13}
L^{(2)}_{E} (t, \h)    =
 - {1 \ov 4} \d_k t_{mn} \d_k t_{mn}  + \ha \d_k t_{mk} \d_n t_{mn}
  + {3 \ov 32} (\d_k \h)^2  + \fo\h  \d_m \d_n t_{mn} \ , 
\ee 
so that integrating out  $\h$ from \rf{13}   we find 
\be \te \la{14}\te 
\bar   L^{(2)}_{E} (t)   = - {1 \ov 4} \d_k t_{mn} \d_k t_{mn}  + \ha \d_k t_{mk} \d_n t_{mn}
  + {1 \ov 6}  \d_m \d_n t_{mn}  \d^{-2} \d_k \d_r t_{kr}  \ .
\ee 
Eq. \rf{14}  is invariant under linearized  reparametrizations 
$\delta t_{mn} = \d_m \xi_n + \d_n \xi_m - \ha \eta_{mn} \d_k \xi_k$  
 with the non-local term   giving the 
required extra terms that were  coming from the variation of $\h$ in the Einstein action. 
Remarkably,  the  quadratic term \rf{14}  has a simple expression in terms of  the 
linearized  Weyl tensor:\foot{Recall that $C^2_{mnkl} = 2(R^2_{mn} - {1\ov 3} R^2)$+ div  and $R_{mn} =- \ha \d^2 h_{mn}  + \d_r \d_{(m} h_{n)r} - \ha \d_m \d_n \h + O(h^2)$.}
\be 
\te 
 \la{112} 
&&\te   \bar  L^{(2)}_{E} (t)  =  \ha C_{mnkl} \d^{-2}   C_{mnkl} = \fo   t_{ab}  P^{ab}_{mn} \d^2 t^{mn}  \ , \qquad \\
&&\te
P^{ab}_{mn} =  P^a_{(m } P^b_{n)}  - {1\ov 3} P^{ab} P_{mn} \ , \qquad 
P_{mn} = \eta_{mn} - {\d_m \d_n\ov \d^2}  \ , \la{113}
\ee
where $P^{ab}_{mn} $ is  the traceless transverse rank 2  projector.\foot{Thus  a 
non-local  redefinition  $t_{mn} \to (\d^2)^{-1/2}  t_{mn}$ 
relates the  quadratic  term in the Weyl theory  to the   quadratic term in the Einstein action with 
the trace $\h$ integrated out. One may wonder if \rf{112} may  have a non-linear   generalization. 
Introducing an auxiliary  tensor $a_{mnkl}$  (with symmetries of the Weyl tensor)  we may consider \rf{112} as 
 a result of integrating out $a_{mnkl}$   in the  Lagrangian  
 $
 L(a)= \ha a_{mnkl} \d^2 a_{mnkl}  -  a_{mnkl}  C_{mnkl}  . 
$
It may first seem  that such a  Lagrangian  could  have   a straightforward non-linear  generalization  given 
that there  exists    a Weyl-covariant generalization of the $\d^2$ operator  acting on rank 4 tensor
 constructed in \ci{erd}.
    However, the action 
    $\int d^4 x  \sqrt g \big[  \ha a^{mnkl} (\nabla^2  + ...) a_{mnkl}  -  a^{mnkl}  C_{mnkl}  \big]$
    can not be  made   Weyl-invariant: 
    if $g_{mn}$   has Weyl weight 1, $C^m_{\ nkl}$   has weight 0  and   
    $a_{mnkl}$   has  weight  3,  then the   first term is  invariant  but the second 
    is not. Thus, in contrast with the $C^2$ action, the $C (\nabla^2 + ...)^{-1}  C$  type 
     action will not be Weyl-invariant.
    }

Solving for $\h$    at the cubic level  
 gives  
 \be   \bar  L^{(3)}_{E} (t)  \te =   X_3 (t)  + {1\ov 3}   X_1 (t)  \d^{-2}  X_2(t)  \ , \qquad  \qquad  X_n(t) \equiv X_n(t, \h=0) \ .  \la{610}
 \ee 
 Similarly, one    can  find  the quartic term $\bar  L^{(4)}_{E} (t) $. 
 The resulting Lagrangian  \rf{aa3}  is   thus non-local at each  order in $t_{mn}$. 
 %
%
It can be simplified by  choosing the transverse gauge
 \be \la{611} 
\d_m t_{mn} =0 \ , \ \ \ \ \ \  \ \  \bar X_n \equiv   X_n(t) \Big|_{\d_k t_{km}=0}  
\ee
and thus fixing the remaining  gauge  symmetry. 
 Then   we get from \rf{a2}--\rf{aa4} (dropping  total derivative terms in $\bar X_3$ and $\bar X_4$) 
 \be \te 
&&\bar X_1 =  0 \ , \ \ \ \qquad 
\te   \bar  X_2 =  {3 \ov 4} \d_k t_{mn} \d_k t_{mn}  - \ha \d_k t_{mn} \d_n t_{mk}   + t_{mn} \del^2 t_{mn}\la{644}  \ ,\\
   && \te 
   \bar X_3 = - \fo t_{ab} \d_a t_{mn}\d_b t_{mn}   + t_{ab} \d_a t_{mn}\d_n t_{mb} 
   - \ha t_{ab} \d_n t_{ma}\d_n t_{mb}  + \ha t_{ab} \d_m t_{na}\d_n t_{mb} 
    \ , \la{733} \\
   &&\te  \bar X_4 =- {1 \ov 16} t_{mn} t_{mn}  ( \d_r t_{ab} \d_r t_{ab}  -2 \d_r t_{ab} \d_b t_{ar})
     \no \\ &&\qquad\quad \te 
    + \ha  t_{ab} t_{cb}  ( \ha \d_a t_{mn} \d_c t_{mn} - \d_n t_{ar}  \d_r t_{cn}  -2 \d_n t_{ar}  \d_c t_{nr} 
  + \d_r t_{na}  \d_r t_{nc} ) \la{633} \\ &&\te 
    + \ha t_{ar} t_{bd}  ( \ha \d_k  t_{ ab} \d_k t_{rd} -  \ha \d_k  t_{ ar} \d_k t_{bd} 
    +  \d_b  t_{ka} \d_d t_{kr} - \d_b  t_{ka} \d_r t_{kd} -2 \d_b  t_{ka} \d_k t_{dr} + \d_k  t_{ar} \d_d t_{bk} )
    \no 
\ee 
Thus    
\be \te
\te \bar  L_E (t) 
=  - {1 \ov 4} \d_k t_{mn} \d_k t_{mn}  + \bar X_3(t)  + \bar X_4(t)  +   Y_4 \ , \ \ \ 
 \  \ \ \   \qquad  Y_4 = {1\ov 6}   \bar X_2 \d^{-2}\bar  X_2  \ ,
  \la{166}   \ee
  i.e.  the non-local  contribution $Y_4$  arising   from integrating out $\h$ starts  at four-point order. 
 
  The resulting three-graviton amplitude is   given by $ \bar X_3(t)$  while the non-local  $Y_4$ 
  contribution to the  four-graviton amplitude may be  represented as\foot{Here we  drop terms with $\d^2 t_{mn}$ as  
  graviton legs  are taken to be on shell. This is justified as long as this  quartic vertex is not inserted in a higher-point scattering 
  amplitude.}
\be \la{566} \te 
 Y_4 =   \te  {1\ov 6} 
   \Big[ {3 \ov 8} \d^2 (t_{mn}  t_{mn} )  - {1\ov 2}  \d_k \d_n ( t_{mn}  t_{mk} ) \Big]   \del^{-2}
     \Big[ {3 \ov 8} \d^2 (t_{ab}  t_{ab} )  - {1\ov 2}  \d_r \d_b ( t_{ab}  t_{ar} ) \Big] \no \\
     \te  = {3 \ov 128 } (t_{mn}  t_{mn}) \d^2 (t_{ab}  t_{ab} )  - {1 \ov 16} (t_{mn}  t_{mn})  \d_r \d_k ( t_{ak}  t_{ar} ) 
     + {1 \ov 24}  ( t_{mn}  t_{mk} )  \d_k \d_n \d^{-2}  \d_r \d_b ( t_{ab}  t_{ar} ) \ . 
\ee
One may  wonder if  this  non-local  $\h$-exchange term 
 should be contributing to the  graviton S matrix    given the 
``unphysical" nature of the trace   field (e.g.  the ``wrong" sign of its  kinetic term in \rf{13}  and  its 
pure gauge role  on shell). 
Also,   given that the action \rf{166}    leads to the same   three-graviton amplitude 
there     should   be no  change to the   graviton S matrix constructed 
according to the BCFW  \ci{bcfw} 
 prescription   where it is determined by unitarity just from   
the three-point graviton  vertex.\foot{
By  unitarity-based arguments  the scattering amplitudes should be cut-constructible.
 The BCFW  representation expresses  the Einstein  four-point 
 S matrix in terms of  $t^3$   physical graviton vertex  (at complex momenta), and thus the trace should not be involved.
 This fixes  the four-graviton vertex in terms of the three-graviton one;
 this is not surprising as the four-vertex should be controlled by gauge invariance.} 
   On  general grounds,   it is the complete four-graviton amplitude
    given by  the $t_{mn}$ exchange part  $\bar X_3 \d^{-2} \bar X_3$ plus  local 
four-vertex $\bar X_4(t) $ plus  the non-local four-vertex $Y_4(t) $   
that should match the four-graviton amplitude in Einstein's theory. 
It is this total amplitude which is    physical and gauge-independent, 
while  the split   between   exchange and contact contributions may depend on
a  choice of  an (on-shell) gauge or a particular choice of polarization tensors.

It may happen that the   non-local term $Y_4$ 
  in \rf{566}    (which by itself is not gauge-invariant) 
   does  not contribute to the S matrix   under  a special    gauge  choice. 
Indeed, as we will  show in   Appendix C,  there exists   a  choice of  graviton polarization tensors
for which  the on-shell    matrix element of  $Y_4$  vanishes. 
 The same also applies    to the matrix element  of  the local four-vertex 
 $\bar X_4 $  so that, for this choice of polarization tensors, the total four-graviton amplitude is given 
  just by the graviton exchange contribution.\footnote{This parallels similar choices in pure gauge theories, where special choices 
  of polarization vectors set to zero the contribution of the four-gluon contribution to the tree-level four-gluon amplitude.}  It is  under this special 
  choice that the unitarity-based  BCFW construction 
  of the  four-graviton amplitude  from the three-point vertices  applies.

In  general, the  unphysical trace   exchange contribution   should cancel 
some unphysical (time-like, etc) part of the  $t_{mn}$  exchange   contribution as $t_{mn}$ by itself is not a physical graviton.\foot{
Thus the  trace  can not be  simply dropped out  but  
  should be properly  integrated out, especially in loops (where  one should also  
   take into account its coupling to ghosts).}
  Indeed,  for the agreement  with the  physical light-cone  gauge   approach,  where only
  the  physical   graviton modes  are propagating, 
  the contribution of the trace in the full graviton propagator   should  be canceling against  the contributions of 
other unphysical 
modes contained in  the $t_{mn}$ propagator.
For example, if   we consider  the graviton exchange between two  traceless  and conserved 
stress tensors then the result is simply $T_{mn }\d^{-2} T_{mn}$
as the  trace and longitudinal parts of $t_{mn}$  do not contribute. 
However,   if  the trace of $T_{mn}$ 
is non-zero  then 
its  contribution survives   and for consistency with unitarity (i.e. for the absence of unphysical massless poles) 
its  contribution should   cancel against some  part of the  contribution of the $t_{mn}$ exchange.\foot{
For example,  in the minimally coupled scalar theory 
the contribution of the trace is local: 
$\d^m \p \d_m \p \Box^{-1}  \d^n \p \d_n \p = \fo \p^2 \Box \p^2$, so there is  no contradiction with unitarity.}

\subsection{Conformal  off-shell extension of  the  Einstein theory}
 Let us now show that the same  Lagrangian $\bar L_E(t)$  \rf{aa3}   obtained   by  eliminating  $\h$ from the 
 Einstein Lagrangian  can be found in a  closed form from the 
 Weyl-invariant  off-shell  extension of  the  Einstein theory found   by first  introducing  a  Weyl  ``compensator" -- 
 a  conformally coupled   scalar field -- and then solving for it. Namely, let us   replace the Einstein action by 
  \be \la{1e} 
S(g,\p)= S_E(\p^2 g ) =  \int d^4 x \sqrt g \big( R\,  \p^2 +   6\,  \d^m \p \d_m \p\big)  \ , 
\ee
where $\p$ has  an unphysical   (ghost-like)  sign of  the  kinetic term.
The action \rf{1e}  is invariant under $g'_{mn}= \l^2(x)  g_{mn}, \ \ \p'=\l^{-1}(x)\,  \p$. 
For this  theory to  be perturbatively equivalent to the Einstein   theory, i.e. to   have  the same S matrix, 
one    should  assume  that $\p$  has  a non-zero  constant value  in the flat-space vacuum,\foot{This choice of the vacuum breaks Weyl symmetry spontaneously, and thus also breaks  conformal symmetry of the near-flat-space expansion.}
 i.e. that the expansion near the vacuum   values is  defined  by 
$g_{mn}=\eta_{mn} +   h_{mn}  ,  \ \    \p= 1 + \vp $.  

 If we  fix the Weyl gauge   $\vp=0$ 
we get back to the Einstein theory.\foot{Some previous discussions of this  conformal scalar theory  and its equivalence to the 
 Einstein one   appeared  in  \ci{des,Englert:1976ep,tsam,kall}.} 
 Instead, we  may   solve for $\vp$  in terms  of the metric to obtain a  
``conformal off-shell extension" of the Einstein gravity   \ci{frad} -- 
 a theory which   gives  an equivalent  graviton 
 S matrix   but   has  an additional Weyl symmetry off shell (at the expense 
of   having an extra  non-local term in the  action).  
Explicitly, one finds   \ci{frad}:
\be  \te 
&&\te \p(g) = 1  + \vp (g) \ , \qquad  - \nabla^2 \vp +  {1\ov 6} R (1+  \vp) =0 \ , \ \ \ \quad 
 \vp = -   {1\ov 6}  \Delta^{-1}   R  \ , \quad \Delta \equiv -   \nabla^2  +  {1\ov 6} R \ ,  \no\\
&&\
S_c (g)\equiv S(g,\p(g)) = 6    \int d^4 x \sqrt g \, \p(g)  \Delta \p(g) =    \int d^4 x \sqrt g \Big(  \te R - 
{1\ov 6}  R \Delta^{-1}   R \Big) \ . \la{116}
  \ee
The   additional  Weyl  symmetry  $g'_{mn}= \l^2(x)  g_{mn}$
of this action   
 implies that  $S_c$  depends 
only on  traceless   graviton $t_{mn}$
as one  is allowed  to fix the   traceless  gauge on $h_{mn}$  even {\it off-shell}.\foot{One can check explicitly that $\h$-dependence cancels  between 
the two terms in \rf{116}.}  

Expanding \rf{116}   in powers  of $t_{mn}$   one can  see explicitly, using \rf{a1}--\rf{aa4}, that 
the resulting action is equivalent to $\bar S_E= \int d^4 x \bar L_E(t)$ 
 found  in   \rf{13}--\rf{166}  by integrating out $\h$ from the Einstein action. 
This is   of course  not surprising as  starting   with \rf{1e}   and  either 
 gauge-fixing $\vp=0$   and  solving for $\h$  or   first  gauge-fixing $\h=0$  and solving for $\vp$ 
should lead to the  same action for $t_{mn}$.\foot{Let us  note 
 that  
the   contribution   of the second  term in \rf{116}   to the four-graviton 
amplitude should be local.  The matrix element of that term  with on shell gravitons   should not have a pole because 
the residue of this pole is the   product of two on-shell matrix elements of $\varphi \bar X_2$    with $\bar X_2$ in \rf{644} 
 which  is zero.
  In general, one   may   find the S matrix of Einstein's  theory by  evaluating the action 
on a perturbative solution of the Einstein equations $R_{mn}=0$   with $h_{mn} =h^{(in)}_{mn} $ 
boundary condition  where $h^{(in)}_{mn} $    is an on-shell graviton mode. 
Then the  bulk of the Einstein action vanishes and  the tree-level S matrix comes from
the boundary term \ci{kgt}.   As is well known,  the  one-loop  correction to the graviton 
 S matrix is finite as the  UV divergences    vanish on shell  \ci{hv}. 
 If we could apply the same argument directly to the second term in \rf{116}  we would conclude that 
 it should  produce trivial contribution to the 
 graviton S matrix  and thus the S matrix   
 should be, as expected,  the same as of the Einstein theory.  While  leading to the correct conclusion, 
  this logic   has a  formal loophole:   to compute the  generating functional for the S matrix 
 of the   theory \rf{116}  we need,   in general, to solve the non-linear equations  following 
 from \rf{116}   rather than those from the Einstein action. 
 The difference compared to the one-loop counterterm example is that there the $R^2$ terms
  are treated as a perturbation, while  in \rf{116}   both terms should a priori  be treated on an equal footing.} 

\subsection{Higher spin  generalization?}

 Let us now  comment on a possible   extension of this construction  to higher spins. 
 Both  conformally  extended Einstein theory \rf{1e}   and the  $C^2$ Weyl theory  share the same symmetries
--  reparametrizations and Weyl invariance  --   but differ in the number of derivatives in the kinetic term  (two  instead of  four). 
 The Weyl theory admits an extension to  the conformal higher spin (CHS)  theory \ci{Fradkin:1985am,Segal:2002gd} 
 which  is invariant under the conformal higher spin symmetry generalizing both the 
 reparametrizations and  the algebraic  Weyl transformations.
 
 This  suggests,  by analogy,   that there may  exist  an ``off-shell extension"  of  a 
 massless  higher spin theory   with two-derivative 
 Fronsdal kinetic terms  
   that  
 contains
   an  extra  tower  of  
     ghost-like  ``compensator"   fields  making   it  invariant   under the same 
 conformal higher spin symmetry   present in the  higher-derivative  CHS theory.
  Solving for this extra    tower 
 of  fields  should then  give an  analog of the non-local action \rf{116}  having 
 an extra  algebraic 
 higher spin conformal symmetry and thus  depending only on the ``physical"  traceless parts 
 $t_s$ of  the original (double-traceless) Fronsdal fields $\p_s$. 
 
 An equivalent   action (leading to the same S matrix) 
  should  originate upon  explicitly integrating out the trace parts $\h_{s-2}$ of the fields $\p_s$   in 
 the interacting   massless   higher spin  Lagrangian 
 $L=\sum_s  \p_s \d^2 \p_s +  V_3(\p)  + V_4 (\p) + ... \  $.
  The  kinetic term 
 in the  resulting non-local action  depending only on the traceless  fields $t_s$   will  be a   generalization of \rf{112}, 
 i.e.  it  may  be  represented  in terms of the same  linearized Weyl tensors  $C_s \sim \d^s t_s $ 
 as the $C_s^2$ kinetic term in the CHS theory, i.e.  
  (cf. \rf{14})\foot{Note that this construction is different 
  from the  one  in \ci{sv} where  the   higher-spin action  depends only on traceless fields  and  is 
local    but  has a  reduced  gauge invariance  (the divergence  of  the gauge parameters is constrained).
It is also different from the approach of  \ci{fras}  which uses 
 the full  linearised higher-spin curvature tensor rather than  its  Weyl part  and thus 
  does not have   the conformal higher-spin symmetry  and involves
    more field components (unconstrained fields  with non-zero  traces, etc).}
 \be \la{6337}  C_s \Box^{1-s} C_s = t_s \del^2 t_s + ... \ . \ee
 \ar{Using  the known  local  Lorentz-covariant   cubic Fronsdal field  interaction vertex $V_3$    \ci{Manvelyan:2010jr},
  it is possible to  find explicitly  the 
 corresponding non-local    contribution to the four-point vertex  generated by  integrating out 
 the ``trace" fields $\h_{s-2}$ 
 which  should generalize the $Y_4$  term  in \rf{166}. 

While  not directly   related,  
an   extended  higher spin theory  with      conjectured local
 action  starting   with  \rf{577}   that involved  an extra tower of  ``ghost-like"  higher-spin 
fields $\psi_j$   appears to  resemble
 such a conformal extension  of the massless  higher spin theory. 
The  non-localities   found  upon   eliminating  the extra fields  may  look somewhat 
analogous  to the  ones appearing in the higher spin 
generalization of \rf{166},\rf{566},\rf{116}. 
The ghost-like   nature  of the additional fields $\psi_j$ suggests, that like the 
trace  fields $\h_{s-2}$  or  conformal  compensator fields of the conformal off-shell extension 
they should  not appear as  asymptotic states in the S  matrix.

More explicitly,  one   could  speculate that the extended  local   theory discussed in section~5 may be  a generalization  of 
the action \rf{1e}  } expanded near flat space vacuum  ($g=\eta+h, \ \p= 1 + \vp$) before  fixing the Weyl symmetry, 
i.e. of  $L(h, \vp) = \sqrt g \big[ R +  R \vp (2+\vp) + 6 (\del \vp)^2  \big]$, 
  with $\psi_j$ in \rf{577} 
being the counterparts of  the ghost-like  field $\vp$. 
Having  introduced the tower of $\psi_j$  one could    discover   that  the resulting action 
 has  a hidden infinite dimensional symmetry 
 (generalizing the Weyl symmetry of $L(h, \vp)$ given above).
 This  symmetry  may be  the conformal higher spin symmetry containing transformations 
  that act ``non-diagonally"    on the infinite set of   higher spin fields (relating   fields of different  spins).

\ar{ While this speculative scenario  has an obvious flaw in that the conformal  off-shell extension   is not supposed to change   the  physical S matrix 
 while the  non-local quartic terms that should be added to the minimal Lorentz-covariant higher spin   action 
  do  contribute to the S matrix, it  still has appealing features. 
 For example, the  presence of a   hidden   infinite dimensional 
conformal higher spin symmetry  may   provide  an explanation  of   why   the resulting   S matrix  may  be trivial, 
as it is the case in the conformal higher   spin  theory   \cite{Joung:2015eny,Beccaria:2016syk}.\foot{By S matrix here we mean the 
  one for the  original   physical massless fields $\p_j$  and not $\psi_j$ -- like $\vp$ in the above example of  $L(h, \vp) $ 
  above they  may not appear  as asymptotic states.}
 }

\section{Concluding remarks} 

Gauge invariance of the S matrix is a powerful tool for 
 constraining the underlying  Lagrangian. 
  Locality of the Lagrangian is reflected in  the S matrix   having 
   poles whose residues are products of lower-point amplitudes.
   This 
    circumvents the  need  for finding nonlinear deformations of the 
    symmetry transformations simultaneously with the construction of quartic and higher-point 
    interaction  vertices   in the Noether procedure. 
  
Using this S matrix  based  approach we  have shown that there exist 
local quartic Lagrangians such that the four-point S-matrix elements of 
three spin-0 particles and either one spin-2 or one spin-4 particles are 
 gauge invariant in   \ar{a higher-spin theory    containing a single tower of massless higher-spin fields
 with local Lorentz-covariant cubic interactions.}
 As in \ci{Ponomarev:2016jqk}   we  have  used a specific  choice of the three-point coupling 
 coefficients  \rf{29} 
 found in \ci{Metsaev:1991mt},  but  our main conclusions  
 should hold for a generic choice of these couplings.
  
 \ar{ For  spins higher than four  this  is no longer possible   without adding  non-local quartic   vertices.  
  We computed a minimal set of non-local quartic 
   terms demanded by gauge invariance and   proposed 
    that they may be eliminated by introducing an additional tower of ghost-like 
   massless  higher-spin fields.
For this  procedure  to work one 
also needs, in particular, 
 a quartic nonlocal interaction of four spin-0 fields and two spin-0  fields with  two 
 higher-spin fields. The former are such that they  appear  to completely 
 cancel the pole terms in the exchange part of the $0000$   amplitude  
 suggesting that it may,  in fact,   vanish identically.  The same  may  apply also to  other amplitudes.} 
  
  While the presence of this  non-local  four spin-0 term 
cannot be tested by  the gauge-invariance considerations, this may be possible  for the terms 
with two higher-spin fields. 
It would be interesting  
 to study the constraints imposed on the  higher-spin Lagrangian 
by the gauge invariance of the $00j_1 j_2$ amplitude  and 
check whether  the terms in \eqref{59} are both necessary and sufficient 
for the gauge invariance. A positive result would be a strong indication that the introduction  of the 
extra ``ghost-like"  fields  to make the action local is indeed a   natural step 
and that the complete S matrix  of the resulting  theory may,  in fact,  be trivial.

It would be interesting to  extend the  discussion of this paper 
to  
higher-point  amplitudes and extract from them the 
corresponding higher-point Lagrangian terms. 
There are two possible approaches that one may use (which should be equivalent up 
to local field redefinitions).
Assuming that one  have found a nonlocal Lagrangian up to  terms with $n$ fields one may construct the exchange
part  of the $(n+1)$-point amplitude and then   fix an $(n+1)$-field term in the  Lagrangian  needed  to restore gauge 
invariance of the S matrix.  It  may happen  that the resulting   non-local  terms may be  replaced by local 
terms  by introducing a suitable set of auxiliary  ghost-like fields. 
Alternatively, one may start with a   local  cubic  plus  quartic Lagrangian of the  extended   theory 
  and analyze only 
the four-point amplitude but with {\em any} external legs, including the auxiliary ghost-like fields. Gauge invariance will 
demand again the presence of a non-local quartic vertices which one  
 may  then convert  again into local interactions by introducing  further  auxiliary fields, etc.   
Integrating out the first $n$  towers  of auxiliary fields   should reproduce the Lagrangian obtained 
in  the first approach,   up  to 
terms with $n+3$ fields; it will  also contain nonlocal interactions of the higher towers of   auxiliary fields.




With a motivation to understand possible types of  non-localities in higher spin  actions 
we have also discussed the conformal off-shell extension of Einstein's gravity and 
showed that its perturbative action  is  equivalent to  the nonlocal action obtained by
 integrating out the graviton trace in the standard Einstein Lagrangian.
\ar{ By analogy, we conjectured  the existence of a conformal off-shell extension of massless higher spin theory 
 containing, in addition to the original tower of the Fronsdal fields,  also a tower of  ghost-like compensator  fields 
  and    noted a  certain  similarity  to  an    extended  local  action  that is  suggested   by 
   S matrix considerations.
   We conjectured  that the latter  may have 
  the    same infinite-dimensional symmetry as the conformal higher-spin theory and  may 
     thus   have a trivial S matrix.} 
Assuming the conformal  extension of the Fronsdal massless  higher spin theory  may indeed exist, 
  it may also provide a link to the 
massless   higher spin theory in AdS space by choosing a
 different vacuum expansion point:  AdS  instead of the  flat space. 


%

\iffa 
The construction of an action for the minimal higher-spin theory 
in AdS or dS space, capturing the physics of Vasiliev's equations, remains an 
open problem. A possible approach is to start with the conformal higher-spin theory in these spaces, 
for which an action is known, 
spontaneously break conformal symmetry while 
also integrating out the conformal compensator fields.
 Our discussion of the conformal extension of Einstein's gravity suggests that the S matrix of the resulting theory -- or rather its Witten diagrams -- should be the same as 
those of the minimal higher-spin theory action with the traces of higher-spin fields  integrated out.
By taking its flat-space limit it may be possible to construct an action for the flat space 
higher-spin theory. It should be interesting to see 
how/if the triviality of the scattering matrix emerges in this approach.

Similarly, the action for conformal higher-spin theories in flat space space may be obtained 
by coupling these fields with one free scalar and select 
the singular term from integrating it out at one-loop level. 
Our discussion suggests, similarly to the AdS case, 
a possible approach to constructing a nonlocal minimal 
higher-spin theory in flat space is to integrate out conformal compensator fields in this action. 
\fi

\section*{Acknowledgments}

We would like to thank  R. Metsaev, D. Ponomarev, E. Skvortsov  and    M. Taronna
 for  very useful  discussions and comments on the draft. 
 We are grateful to M. Taronna    for sharing with us a draft of   his forthcoming  paper \ci{tar17}. 
 The work of RR was supported by the DOE grant DE-SC0013699. 
 The work of AAT was supported by the ERC Advanced grant no. 290456, the STFC Consolidated grant ST/L00044X/1,  
   the Australian Research Council, project DP140103925, 
 and by the Russian Science Foundation grant 14-42-00047 at Lebedev Institute.


\appendix

\section{Contact term contribution to the $000j$  scattering amplitude  
\label{contact_term}  }

The contact term in the $000j$ amplitude following from the quartic Lagrangian \eqref{21} is:
\be
{\cal A}^{ct} &=& 
       i\,\rC_{j}{}_0 (p_1, p_2, p_3)\phi_j(p_4, (2i p_3)^j)+i\,\rC_{j}{}_0 (p_2, p_1, p_3)\phi_j(p_4, (2i p_3)^j)\no 
\\
&+&i\,\rC_{j}{}_0 (p_1, p_3, p_2)\phi_j(p_4, (2i p_2)^j)+i\,\rC_{j}{}_0 (p_3, p_1, p_2)\phi_j(p_4, (2i p_2)^j)
\cr
&+&i\,\rC_{j}{}_0 (p_2, p_3, p_1)\phi_j(p_4, (2i p_1)^j)+i\,\rC_{j}{}_0 (p_3, p_2, p_1)\phi_j(p_4, (2i p_1)^j)
\cr
&+&i \,(\rC_{j}{}_{j/2} (p_1, p_2, p_3)+\rC_{j}{}_{j/2} (p_1, p_3, p_2))\phi_j(p_4, (2i p_2)^{j/2}, (2i p_3)^{j/2})
\cr
&+&i \,(\rC_{j}{}_{j/2} (p_2, p_3, p_1)+\rC_{j}{}_{j/2} (p_2, p_1, p_3))\phi_j(p_4, (2i p_1)^{j/2}, (2i p_3)^{j/2})
\cr
&+&i \,(\rC_{j}{}_{j/2} (p_3, p_1, p_2)+\rC_{j}{}_{j/2} (p_3, p_2, p_1))\phi_j(p_4, (2i p_1)^{j/2}, (2i p_2)^{j/2})
\label{contact}
\\
&+&i\sum_{k=1}^{j/2-1} \big(\rC_{j}{}_k (p_1,  p_2, p_3)\phi_j(p_4, (2i p_2)^k,(2i p_3)^{j-k})+\rC_{j}{}_k (p_1,  p_3, p_2)\phi_j(p_4, (2i p_3)^k,(2i p_2)^{j-k}) \big)
\cr
&+&i\sum_{k=1}^{j/2-1} \big(\rC_{j}{}_k (p_2,  p_3, p_1)\phi_j(p_4, (2i p_3)^k,(2i p_1)^{j-k})+\rC_{j}{}_k (p_2,  p_1, p_3)\phi_j(p_4, (2i p_1)^k,(2i p_3)^{j-k}) \big)
\cr
&+&i\sum_{k=1}^{j/2-1} \big(\rC_{j}{}_k (p_3,  p_1, p_2)\phi_j(p_4, (2i p_1)^k,(2i p_2)^{j-k})+\rC_{j}{}_k (p_3,  p_2, p_1)\phi_j(p_4, (2i p_2)^k,(2i p_1)^{j-k}) \big)
\nonumber
\ee
To study the cancellation of its gauge variation against that of the exchange part of the amplitude 
 it is important to choose an independent basis of  contractions of the gauge parameter. 
This can be facilitated  by  writing  the amplitude in terms  of 
an independent basis of   contractions of 
the spin-$j$ polarization tensor with momenta.  Eliminating $p_1$ from these contractions we find:
\be\no 
&&{\cal A}^{ct} = i {\cal B}_2  \phi_j(p_4, (2i p_2)^j) +  i {\cal B}_3  \phi_j(p_4, (2i p_3)^j)+i {\cal B}_{23}  \phi_j(p_4, (2i p_2)^{j/2}, (2i p_3)^{j/2})
\\
&&\   +\,  i\sum_{n=1}^{j/2-1} {\cal D}_{23,n} \phi_j(p_4, (2i p_2)^n, (2ip_3)^{j-n})+ i\sum_{n=1}^{j/2-1} {\cal D}_{32,n} \phi_j(p_4, (2i p_3)^n, (2ip_2)^{j-n})
\ 
\label{ActGeneral}
\ee
with the coefficients  given   by 
\be
{\cal B}_2&=& 
\,\rC_{j}{}_0 (p_1, p_3, p_2)+\rC_{j}{}_0 (p_3, p_1, p_2)\ 
+\ 
\, \rC_{j}{}_0 (p_2, p_3, p_1)+\rC_{j}{}_0 (p_3, p_2, p_1) 
\cr
&+&(-1)^{j/2} \,(\rC_{j}{}_{j/2} (p_3, p_1, p_2)+\rC_{j}{}_{j/2} (p_3, p_2, p_1))
\cr
&+& \sum_{k=1}^{j/2-1}(-1)^k  \rC_{j}{}_k (p_3,  p_1, p_2) 
+
\sum_{k=1}^{j/2-1}(-1)^k \rC_{j}{}_k (p_3,  p_2, p_1)\ , 
\label{B2}
\ee
\be
{\cal B}_3 &=& 
  \,\rC_{j}{}_0 (p_1, p_2, p_3)+\rC_{j}{}_0 (p_2, p_1, p_3)\ 
  + \
\, \rC_{j}{}_0 (p_2, p_3, p_1)+\rC_{j}{}_0 (p_3, p_2, p_1)
\cr
&+&(-1)^{j/2}\,(\rC_{j}{}_{j/2} (p_2, p_3, p_1)+\rC_{j}{}_{j/2} (p_2, p_1, p_3)) 
\cr
&+& \sum_{k=1}^{j/2-1} (-1)^k \rC_{j}{}_k (p_2,  p_3, p_1)
+
\sum_{k=1}^{j/2-1} (-1)^k \rC_{j}{}_k (p_2,  p_1, p_3)
\la{a4}  \ ,  \ee
\be
{\cal B}_{23} (p_1, p_2, p_3)&=& 
 \rC_{j}{}_{j/2} (p_1, p_2, p_3)+\rC_{j}{}_{j/2} (p_1, p_3, p_2)
+
\,\big(\rC_{j}{}_0 (p_2, p_3, p_1)+\rC_{j}{}_0 (p_3, p_2, p_1)\big) C_j^{j/2}
\cr
&+&(-1)^{j/2} \,(\rC_{j}{}_{j/2} (p_2, p_3, p_1)+\rC_{j}{}_{j/2} (p_2, p_1, p_3))  
\cr
&+&(-1)^{j/2} \,(\rC_{j}{}_{j/2} (p_3, p_1, p_2)+\rC_{j}{}_{j/2} (p_3, p_2, p_1))   
\cr
&+& \sum_{k=1}^{j/2-1} (-1)^k \rC_{j}{}_k (p_2,  p_3, p_1)C_{j-k}^{j/2-k} 
+
\sum_{k=1}^{j/2-1}(-1)^k \rC_{j}{}_k (p_3,  p_2, p_1) C_{j-k}^{j/2-k} \ , 
\la{a5}  
\ee
\be
{\cal D}_{23,n}(p_1,p_2, p_3) &=&
 \,\big(\rC_{j}{}_0 (p_2, p_3, p_1)+\rC_{j}{}_0 (p_3, p_2, p_1)\big)C_j^n 
 \cr
&+&
(-1)^{j/2}  \,(\rC_{j}{}_{j/2} (p_2, p_3, p_1)+\rC_{j}{}_{j/2} (p_2, p_1, p_3)) C_{j/2}^n  
\cr
&+&   \rC_{j}{}_n (p_1,  p_2, p_3) 
+ 
\sum_{k=1}^{n}(-1)^k \rC_{j}{}_k (p_3,  p_2, p_1)  C_{j-k}^{n-k}
\cr
&+&  
 \sum_{k=n}^{j/2-1}(-1)^k \rC_{j}{}_k (p_2,  p_1, p_3)C_k^n
+\sum_{k=1}^{j/2-1} (-1)^k \rC_{j}{}_k (p_2,  p_3, p_1)C_{j-k}^{j-k-n}\  
\la{a6} 
\ee
\be
{\cal D}_{32,n} (p_1,p_2, p_3)&=& 
\,\big(\rC_{j}{}_0 (p_2, p_3, p_1)+\rC_{j}{}_0 (p_3, p_2, p_1)\big) C_j^{j-n}  
\cr
&+&
(-1)^{j/2} \,(\rC_{j}{}_{j/2} (p_3, p_1, p_2)+\rC_{j}{}_{j/2} (p_3, p_2, p_1))) C_{j/2}^n  
\cr
&+& \rC_{j}{}_n (p_1,  p_3, p_2)  
+
\sum_{k=1}^{n} (-1)^k \rC_{j}{}_k (p_2,  p_3, p_1) C_{j-k}^{n-k}  
\cr
&+& \sum_{k=n}^{j/2-1}(-1)^k  \rC_{j}{}_k (p_3,  p_1, p_2)C_k^n 
+
 \sum_{k=1}^{j/2-1}(-1)^k \rC_{j}{}_k (p_3,  p_2, p_1)C_{j-k}^{j-k-n}\ 
 \label{a7}
\ee
It is then  straightforward to compute the  variation of ${\cal A}^{ct}$ in \rf{contact} under the gauge transformation \eqref{k_gt}.
 The resulting expression was  given 
in eq.~\eqref{gtContactFinal} in the main text.

\section{Soft momentum expansion of  massless higher spin  amplitudes\\
and gauge invariance constraints
\label{B} }

Soft momentum limits  of scattering amplitudes 
are known   to contain information about  symmetries of a theory.
In this Appendix  we   shall analyze the soft-momentum 
limit of a  massless  higher-spin theory 
with  generic  three-spin couplings given by  vertices $V_{j_1,j_2,j_3}$.  
We shall  assume that the theory is local, {\it i.e.} that all  poles  in momentum variables   appearing in the 
 (integrand of) 
  scattering amplitudes  may only come   from  on-shell propagators of particles present  in the 
 original  action. 
 %
 We shall moreover assume that the three-field terms are not gauge invariant off shell (i.e. they are 
 not constructed out of tree field strengths) but rather  are invariant only on shell (i.e. up to terms proportional to 
 the free equations of motion).

We shall generalize the discussion in \cite{Bern:2014vva}, which itself generalizes that of \cite{Low:1958sn}.
We shall restrict 
 consideration  to  the leading order of soft momentum expansion, 
 which extends the considerations in \cite{Weinberg:1964ew} to effectively
  arbitrary couplings of higher-spin fields.
  Similar analysis  was carried out earlier  in \ci{Taronna:2011kt}  with  equivalent conclusions.

For notational simplicity we  shall  use a  somewhat   symbolic notation 
not including  the polarization tensor  factors  until the discussion of gauge transformation 
of the amplitude. 

\subsection{Expansion of the $0...0j$ amplitude}

\begin{figure}[t]
      \centering
      \includegraphics[scale=0.75]{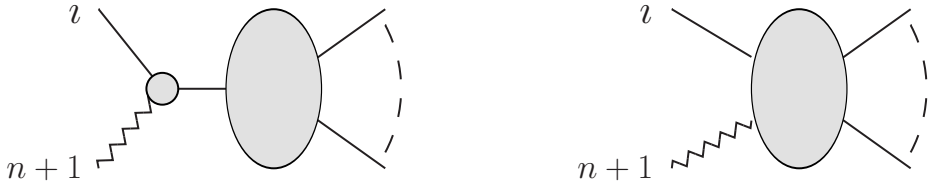}
      \caption[a]{\small The two types of contributions to a soft limit of an amplitude with $n+1$ particles. 
       The first  has a pole when $p_{n+1}\rightarrow 0$ while the second  does not. 
        The zig-zag line represents a particle of spin $j$  whose  momentum is taken to be  soft;
         straight lines represent particles of spin $j_i$   ($i=1, 2, ..., n$). 
      \label{factorization} }
\end{figure}

We shall start with  an amplitude for  $n$ spin-0 particles with  momenta $p_1,\dots,p_n$ and one spin-$j$ particle with 
momentum $p_{n+1}\equiv q$  which will be taken to be small.  In the $q\to 0$  limit 
 there are two  contributions to  this amplitude  shown in fig.~\ref{factorization}. Isolating the pole part 
 from the regular part (denoted by $N$   below)
 and using the expression for the $(0j_1 j_2)$    cubic vertex in \rf{8}   we  get 
\be
\cA^{\mu_1\dots \mu_j}(p_1,\dots,p_n,q) &\sim & \sum_{i}\sum_{\tsj_i} 2^{j-1}\frac{p_i^{\mu_1}\dots p_i^{\mu_j}}{q\cdot p_i}
((p_i-q)\cdot \partial_u)^{\tsj_i}\  W_{\tsj_i}(p_i+q, \partial_{u'}) P_{\tsj_i}(u,u')
\cr
&&
+\ N^{\mu_1\dots \mu_j}( p_1,\dots,p_n,q)\la{b1} 
\\
&=& \sum_{i}\sum_{\tsj_i} 2^{j-1}\frac{p_i^{\mu_1}\dots p_i^{\mu_j}}{q\cdot p_i}
 \ \tsj_i! \ W_{\tsj_i}(p_i+q, \partial_{u'}) P_{\tsj_i}(p_i-q,u')
\cr
&&
+\ N^{\mu_1\dots \mu_j}( p_1,\dots,p_n,q) \ .\la{b2} 
\ee
Here   we used that  $(p_i +q)^2 = 2 q\cdot p_i  + q^2  \to  2 q\cdot p_i $   in the $q\to 0$ limit. 
The free  indices $\mu_1,\dots,\mu_j$ are to be contracted 
with the spin-$j$ polarization tensor of the  external soft field.
The derivatives with respect to $u$ and $u'$  represent  the contraction 
of the three-point vertex and  the function  $W$ through the
 transverse-traceless projector $P_{j'} (u,u') $ in the propagator of an  internal spin-$j'$ field;
  $u,u'$ should be set  to zero once the derivatives $\partial_u$  and $\partial_{u'}$ are evaluated
   (see, e.g., \cite{Ponomarev:2016jqk} for details on the Feynman rules of  Fronsdal higher-spin fields).


  The factor 
  $W_{\tsj_i}$  (represented by the  right  blob in the first diagram in fig.~\ref{factorization})
   is a Green's function with all 
  but the $i$-th leg (carrying  momentum $p_i+q$) being  on shell.  When $q=0$ it becomes 
an $n$-point amplitude once it  is   contracted with  a polarization tensor. 
Since for $q=0$  it is a scattering  amplitude
 it   should be  gauge-invariant. Thus  its  contraction 
 with  a polarization vector containing  a factor of the corresponding momentum
    should vanish. Therefore,  $(W_{\tsj_i})_{q\to 0}= W_{\tsj_i}(p_i, \partial_{u'})  $ must obey the following relations
(for all values of $\tsj_i$ and $k$)
\be
&& W_{\tsj_i}(p_i, \partial_{u'}) P_{\tsj_i}(p_i,u') = 0 ~, ~ ~~\qquad\qquad \qquad \tsj_i\ne 0  \ , 
\\
&& W_{\tsj_i}(p_i, \partial_{u'}) (p_i\cdot u')^ k P_{\tsj_i-k}(p_i,u') = 0~, ~ ~~ \qquad k=1,\dots \tsj_i \ . 
\label{gi_constraints}
\ee
At the  same time,  the  gauge invariance of the full amplitude \rf{b1}  with respect to the transformation of 
the spin $j$  polarization tensor  requires that 
\be
 q_{\mu_j} \cA^{\mu_1\dots \mu_j}(p_1,\dots,p_n,q)=0 \ , 
\ee
or,  equivalently,  
\be
 \sum_{i}\sum_{\tsj_i} {p_i^{\mu_1}\dots p_i^{\mu_{j-1}}}
\ \tsj_i!\ W_{\tsj_i}(p_i+q, \partial_{u'}) P_{\tsj_i}(p_i-q,u')
+q_{\mu_j} N^{\mu_1\dots \mu_j}( p_1,\dots,p_n,q)=0 \  . \ \ 
\label{b6}
\ee
This relation  should hold for any $q$, e.g., order by order in a small $q$ expansion. 
Below   we shall   focus on  the leading order in   its  small-$q$ expansion.

The leading   ${\cal O}(q^0)$  term   obtained by setting $q=0$  in  \rf{b6}   gives 
\be
 \sum_{i}\sum_{\tsj_i} {p_i^{\mu_1}\dots p_i^{\mu_{j-1}}}
\ \tsj_i!\ W_{\tsj_i}(p_i, \partial_{u'}) P_{\tsj_i}(p_i,u')=0 \ . \la{b7}
\ee
Here we used the assumption   of locality to drop  the $N$-term in \rf{b6}   which should not  have poles in $q$. 
Since  here  $W_{\tsj_i}$ (taken at $q=0$)    is an on-shell amplitude, it   should   obey the  gauge-invariance 
constraints \eqref{gi_constraints}  which   imply that only the terms with $\tsj_i=0$ are non-vanishing. 
Then the  surviving  $W_0$ factor  becomes   simply the same as  the
 scattering amplitude $ \cA^{0...0}(p_1,\dots,p_n)$  of $n$ spin-$0$ fields, i.e.   we get 
(for any $j$) 
\be
 \cA^{0...0}(p_1,\dots,p_n) \sum_{i} p_i^{\mu_1}\dots p_i^{\mu_{j-1}}=0 \ .  
\ee
As  the  sum of  products of momenta does not, in general, vanish   if $j >2$  we conclude that 
\be
\cA^{0...0}(p_1,\dots,p_n) = 0 \ , 
\label{b99}
\ee
i.e. if one starts with a {\it local}  action then  the scalar  scattering amplitude should vanish, 
in agreement  with  \cite{Weinberg:1964ew}.

Note that this  conclusion  is consistent with the cancellation of the poles 
of the $0000$ amplitude observed in section~5  where we  introduced  a local action 
involving extra fields  $\psi_j$.\foot{While the couplings of these  additional fields  $\psi_j$ are somewhat different from those of the minimal set of fields, their momentum dependence  is the same; therefore, the analysis  of  this Appendix 
should  hold   also  in the presence of these  additional fields.}
%

\subsection{Expansion of the $j_1...j_n j$ amplitude}

Next, let   us   generalize the above   discussion  and 
  consider  the consequences of  gauge invariance  for an 
amplitude with $n$  particles of  generic spins $j_1,\dots, j_n$ (with momenta $p_1,\dots,p_n$)  and
 an additional particle  of spin $j$  with   soft momentum $p_{n+1}=q$. 
 We will  again assume  that our starting point is a local  interacting Lagrangian. 
 As in the $j_1=...=j_n=0$ case   analysed  above, 
 the  $q\to 0$ limit  of this  amplitude  then  has  a  singular 
  and a  regular  contributions shown in fig.~\ref{factorization}:
\be
\cA^{\mu_1\dots \mu_j}(p_{1},\dots,p_{n},q) &\sim & \sum_{i}\sum_{\tsj_i} 
V^{\mu_1\dots \mu_j}_{j,j_i,\tsj_i}(q, p_i, \partial_u) \ \frac{P_{\tsj_i}(u,u')}{2q\cdot p_i}\ 
W_{\tsj_i}(p_i+q, \partial_{u'}) 
\cr
&&
+\ N^{\mu_1\dots \mu_j}(p_1,\dots,p_n,q) \ . 
\label{generalform_all}
\ee
Here we explicitly indicated  only Lorentz  indices  that should be   contracted   with spin $j$ polarization tensor.
As in \rf{b1}, 
$W_{\tsj_i}$ is a Green's function with all but the $i$-th leg (with momentum $q+p_i$) 
 being  on shell. For  $q\to 0$ it becomes 
an $n$-point amplitude once it acts on a polarization tensor, 
and  thus   it should  obey the constraints \eqref{gi_constraints}.

The covariant three-point vertices  $V^{\mu_1\dots \mu_j}_{j,j_i,\tsj_i}(q, p_i, \partial_u)$ 
  can be found  in \cite{Manvelyan:2010jr}. 
In the present  case (relevant   for the  discussion of the first diagram  in fig. \ref{factorization}) 
 the cubic vertex    may   be written as 
\be
V_{j,j_i,\tsj_i}(q,p_i,p_{i'} ; u_q; \partial_u) 
&\equiv&
u_q^{\mu_1}\dots u_q^{\mu_j}  V^{\mu_1\dots \mu_{j}}_{j,j_i,\tsj_i}(q,p_i,p_{i'}; \partial_u) 
\no \\
&=&c_{j j_i \tsj_i}
\sum_{\stackrel{\alpha+\beta+\gamma=n}{0\le n\le {\rm min}(j, j_i, \tsj_i)}}
\frac{1}{\alpha!\beta!\gamma!}
((q-p_i)\cdot \partial_u)^{\tsj_i-n+\gamma}
\la{b11} \\
&&
(u_q \cdot (p_i-p_{i'}))^{j - n+\alpha} (u_q\cdot  \partial_u)^{\beta}
\phi_{j_i}\big(p_i, (p_{i'}-q)^{j_i-n+\beta},\partial_u^\alpha,u_q^\gamma\big) \ .\no 
\ee
Here we assume that the polarization tensor of the external leg $\phi_i$   is already a part of $V$  
and the argument $u$   corresponds to the internal line  with spin $j'_i$   and momentum $p_{i'}$. 
We also used the notation $\phi_j(a^k, ... ,b^n)$   explained  in \rf{4}. 
The  coefficients $c_{j j_i \tsj_i}$ (fixed in the light-cone gauge approach \ci{Metsaev:1991mt} 
 will be assumed a priori  to be arbitrary. 

The   transformation of $V_{j,j_i,\tsj_i}$ under the spin-$j$ gauge symmetry 
is given by the contraction of one of its free indices with the momentum  $q$, or, alternatively, by replacing one of the vectors $u_q$ 
by  $q$.  Then the only nontrivial contribution 
comes from $q\cdot (p_i-p_{i'})$ term; 
all other terms cancel out because of the on-shell gauge invariance of the three-point vertex   \cite{Manvelyan:2010jr}, i.e. 
\be
q\cdot \partial_{u_q} V_{j,j_i,j_3}(q,p_i,p_{i'}; u_q; \partial_u)  &=& 2 q\cdot p_i \ c_{j j_i \tsj_i}
\sum_{\stackrel{\alpha+\beta+\gamma=n}{0\le n\le {\rm min}(j, j_i, \tsj_i)}}
\frac{1}{\alpha!\beta!\gamma!}
((q-p_i)\cdot \partial_u)^{\tsj_i-n+\gamma}
\no \\
&& \!\!\!\!\!\!\!\!\!\!\!\!
\times (u_q\cdot (p_i-p_{i'}))^{\alpha-1} (u_q\cdot \partial_u)^\beta
\phi_{j_i}(p_i, (p_{i'}-q)^{j_i-n+\beta},\partial_u^\alpha, u_q^\gamma ) 
\cr
&\equiv & 2 q\cdot p_i\  F_{j,j_i,\tsj_i}(q, p_i;u_q; \partial_u)\ . 
\ee
Therefore, the spin-$j$ gauge invariance of the amplitude \eqref{generalform_all} implies  that 
\be
0 &=& q\cdot \partial_{u_q} (u_q^{\mu_1}\dots u_q^{\mu_j} \cA^{\mu_1\dots \mu_j}(p_1,\dots,p_n,q)) 
\no \\
&=& 
\sum_{i}\sum_{\tsj_i}  F_{j,j_i,\tsj_i}(q, p_i;u_q; \partial_u) \ 
W_{\tsj_i}(p_i+q, \partial_{u'}) \  {P_{\tsj_i}(u,u')}
\cr
&&
+\ q\cdot \partial_{u_q}  (u_q^{\mu_1}\dots u_q^{\mu_j}  N^{\mu_1\dots \mu_j}(p_1,\dots,p_n, q)) \ .
\label{b13}
\ee
Here the  second line  is the transformation of the contribution of  the first diagram in fig.~\ref{factorization} and the third 
line represents the transformation of  the contribution of   the second diagram. 

To  determine  the general consequences of  gauge invariance
of the amplitude  we shall expand  \eqref{b13} at small $q$. While there 
may be interesting information contained in the subleading terms (like, {\it e.g.},  subleading soft theorems in 
YM theory \cite{Bern:2014vva}),  here 
we shall restrict  consideration  to the leading  $O(q^0)$ order.


Setting $q=0$ in \eqref{b13} we find  (assuming again the locality of the Lagrangian, i.e. 
the regularity of $N$ in \rf{b13}) 
\be
0 &=& \sum_{i}\sum_{\tsj_i}  F_{j,j_i,\tsj_i}(0, p_i;u_q; \partial_u) \
W_{\tsj_i}(p_i, \partial_{u'}) \ {P_{\tsj_i}(u,u')}
\no \\
&=&
\sum_{i}\sum_{\tsj_i}  c_{j j_i \tsj_i}
\sum_{\stackrel{\alpha+\beta+\gamma=n}{0\le n\le {\rm min}(j, j_i, \tsj_i)}}
\frac{(-1)^{\tsj_i +  j_i-2n+\gamma+\beta}\   2^{j+ j_i  - 2n+\alpha + \beta-1}}{\alpha!\beta!\gamma!}
(p_i\cdot \partial_u)^{\tsj_i-n+\gamma}
\cr
&& \times (u_q \cdot p_i)^{\alpha-1} (u_q\cdot \partial_u)^\beta
\phi_{j_i}(p_i, p_i^{j_i-n+\beta},\partial_u^\alpha,u_q^\gamma) \
W_{\tsj_i}(p_i, \partial_{u'}) \  {P_{\tsj_i}(u,u')}\ . \la{b14}
\ee
The transversality of the polarization tensor and the constraints \eqref{gi_constraints}  imply that 
all the  terms with
$
j_i-n+\beta \ne 0
~~{\rm or}~~
\tsj_i-n+\gamma \ne 0 
$
vanish identically. Therefore, the only nontrivial constraints come from configurations of parameters for which
$
j_i-n+\beta = 0 $ and $
\tsj_i-n+\gamma = 0 
$
are satisfied  at the same time. 
 Since $\beta,\gamma\ge 0$ and $n\le {\rm min}(j, j_i, \tsj_i)$, the solution to these 
constraints is
\be
j_i=\tsj_i \le j 
~~,\quad
n = j_i=\tsj_i
~~,\quad
\beta = \gamma = 0 
~~,\quad
\alpha = n=j_i=\tsj_i \ .
\label{sol}
\ee
Then  all the sums except the sum over the external particles collapse to a single term
and \rf{b14} reduces to 
\be
0 &=&
\sum_{i} c_{j j_i j_i}
\frac{1}{j_i! } (u_q\cdot p_i)^{j-1} 
\phi_{j_i}(p_i, \partial_u^{j_i})\  W_{\tsj_i}(p_i, \partial_{u'}) \ {P_{\tsj_i}(u,u')}
\no \\
&=& \cA^{j_1...j_n} (p_1,\dots, p_n  )  \sum_{i}  c_{j j_i j_i} (u_q\cdot p_i)^{j-1} \ . \la{b17} 
\ee
For $j=2$ 
we get the constraint that $c_{2 j' j'}$  must be the same for all $j'$, i.e. the spin 2 coupling  must  be universal
(then $ c_{2 j' j'}$ factorizes and the momentum conservation sets the sum in \rf{b17} to  zero). 

For  $j >2$  the sum in \rf{b17} cannot vanish  for generic on-shell momenta; 
it follows then that  the  gauge invariance 
requires  that {\it either} all the $j_1...j_n$ amplitudes   
should vanish
\be\la{b18}   \cA^{j_1...j_n} =0 \ , \ee
 {\em or}  we should   have the following constraint  on the three-point coupling constants 
\be\la{b19} 
c_{j j_i j_i} = 0 ~, \ \ \ \ \ \ \ \   ~~ j_i<j \ .
\ee
The latter  condition
  (found  earlier  in \ci{Taronna:2011kt}) 
\rf{b19} means 
that   there should be  no  cubic 
diagonal coupling \rf{b11}  of a spin-$j$ field with  all   smaller  $j_i < j$  spins.
In  our discussion of the $0...0j$  amplitude in the previous subsection we assumed that  $c_{j00} \not=0$ 
and thus  arrived at  \rf{b18}, i.e. \rf{b99}. 

The  3-point coupling constants   found  in the light-cone  approach in \ci{Metsaev:1991mt}
do not satisfy \rf{b19}\foot{We are assuming that the  light-cone  approach  of  \ci{Metsaev:1991mt} 
should correspond to a light-cone gauge fixing in a  Lorentz  and gauge-invariant   action.}
  and
 then  one should  either  relax the  assumption of locality of the  interaction Lagrangian 
or  accept the vanishing of the  scattering amplitudes \rf{b18}.  

\iffa 
 The results  of sections  4 and 5  for $000j$ amplitude with $j \geq 6$ 
  are  consistent with these  conclusions: 
one either needs to add a non-local four-vertex in the action or, if locality can be maintained by introducing an extended set of fields, 
the total amplitude  in the extended  theory (assuming it has a  consistent gauge-invariant formulation) 
should vanish. 
\fi

\section{Vanishing of  the trace   contribution to four-graviton amplitude} 

Below  we sketch the argument for the vanishing of the on-shell matrix element of  the graviton trace  contribution 
in the $\d_m t_{mn}=0$ gauge, i.e.  of $Y_4$ in \rf{566}, for a special  choice of on-shell gauge, i.e. for a special 
choice of the polarization tensors. 


  Let us  for definiteness  consider the  amplitude where the  particles $1$ and $2$ have negative helicity (-2) 
  while the particles $3$ and $4$ have positive helicity  (+2). 
  The main point is  that  using the on-shell gauge invariance  it is possible  
 to  choose the 4 polarization tensors  $ \varepsilon^{\pm} (p)$   in such a way that 
 the following combinations  with some free indices
are simultaneously zero\foot{As in section~6, here   $m,n,...$ are Lorentz indices.}
\be 
 && \varepsilon^-(p_1)_{ac}\, \varepsilon^-(p_2)_{bc}  \, \varepsilon^+(p_3)_{kn} \,\varepsilon^+(p_4)_{mn}  = 0\ , \qquad 
  \varepsilon^-(p_1)_{ac} \,\varepsilon^+(p_3)_{kc}  \, \varepsilon^-(p_2)_{bn} \,\varepsilon^+(p_4)_{mn}  = 0\no \ ,  \\
&&  \varepsilon^-(p_1)_{ac} \,\varepsilon^+(p_4)_{mc}\,   \varepsilon^-(p_2)_{bn}\, \varepsilon^+(p_3)_{kn}  = 0 \ . 
\la{c1}  \ee
Since the matrix element of $Y_4 $  in \rf{566} is given   by  a linear combination of   such products it 
 then vanishes. 

The reason for the    relations \rf{c1}   is the following. 
In the  spinor helicity  notation   the graviton
polarization tensor   may be represented  as (for each helicity choice)
     $   \varepsilon^{\pm} (p) = \varepsilon^{\pm}(p) \otimes\varepsilon^{\pm} (p)$
where $\varepsilon$ is the polarization vector,  
  i.e.    $  \varepsilon^-(p) = |q]\langle p|/[q p] $, \ \ \          $ \varepsilon^+(p) = |q \rangle [p|/\langle q p \rangle   $.  
Here $q = |q\rangle [q|$ is an arbitrary null vector,  which can be chosen 
independently  for each polarization vector; this reflects   the  remaining on-shell  gauge invariance.\footnote{It states that, apart from its own momentum, a polarization vector can be chosen to be orthogonal 
to another arbitrary null vector.}  
Once chosen, these  reference vectors are not changed from graph to graph. 
The  gauge transformation $\varepsilon^\mu \rightarrow p^\mu $
translates into $|q\rangle \mapsto |p\rangle$ and  $|q] \mapsto |p]$.

Choosing $q_1 = q_2$ and $q_3 = q_4$,  the scalar products of the polarization vectors    become 
\be
 && \varepsilon^-(p_1) \cdot \varepsilon^-(p_2) \sim  [q_1 q_2] = [q_1 q_1] = 0\ , 
 \no \\  
 &&\varepsilon^-(p_1)\cdot \varepsilon^+(p_3)\  \varepsilon^-(p_2)\cdot \varepsilon^+(p_4) \sim [q_1 p_3] [q_1 p_4] 
 \langle q_3 p_1 \rangle\langle q_3 p_2 \rangle\ , 
 \no \\
 &&  \varepsilon^-(p_1)\cdot \varepsilon^+(p_4)\  \varepsilon^-(p_2)\cdot \varepsilon^+(p_3) \sim [q_1 p_4] [q_1 p_3]
 \langle q_3 p_1\rangle\langle q_3 p_2\rangle \ . \la{c2}
\ee
Then choosing,  e.g.,  $q_1 = q_2 = p_3$ and $q_3 = q_4 = p_1$ sets to zero the last   two products in \rf{c2} 
and  thus implies  \rf{c1}  and, as a consequence,  the vanishing of the matrix element of $Y_4$.

The  above  choice   simplifies   also  the on-shell 
matrix elements of $\bar X_3$ in \rf{733}  and $\bar X_4$  in \rf{633}  since 
 whenever two  external on shell polarization tensors 
are contracted over (at least) one index they do not contribute  to the amplitude.  For example, 
in  $\bar X_4$ there are always at least two  $t_{mn}$  tensors contracted over one index, so with the above choice of 
the polarization tensors   the  contribution  of this local four-point   vertex  to the amplitude  also  vanishes. 
This mirrors  what happens in gauge theory where a similar  
 choice of  the  reference vectors $q_i$  sets to zero  the contribution of the four-point contact 
term in  the YM  action to the tree-level four-gluon amplitude;  then  the full amplitude is given just 
by the exchange contribution.

\iffa 
 this is  indeed in parallel to AAAA  not
contributing to four-gluon amplitude (though at first sight that looks
rather strange and artificial choice). But reason should be that
four-point couplings are dictated  by gauge invariance.
Hope was to learn something useful for  higher spins. It seems the
message is that we  can basically ignore traces there too -- consider
only traceless
(rather than double-traceless)  fields as they will only  be contributing
to onshell amplitudes. Then the issue is  if these nonlocal terms
that were appearing before are like that Y-term or  they really
matter. Looked the second -- as they were  already in on-shell
amplitude...
\fi



\end{document}